\documentclass[oneside]{amsart}
\usepackage{color,pstricks,caption}

\newtheorem{example}{Example}[section]

\newtheorem{theorem}{Theorem}

\newtheorem{lemma}[example]{Lemma}

\newcommand{\Proof}{\medskip\noindent\emph{Proof: }}
\newcommand{\cqfd}{\hfill $\Box$\medskip}
\newcommand{\M}{\mathcal M}
\newcommand{\boldB}{\boldsymbol{B}}
\newcommand{\boldn}{\boldsymbol{n}}

\newcommand{\wt}{{wt}}
\newcommand{\wh}{\widehat{wt}}
\newcommand{\unwt}{{{\hat{w}}^{\circ}}}

\newcommand{\hh}{\hat h}
\newcommand{\ha}{\hat a}
\newcommand{\hb}{\hat b}
\newcommand{\hx}{\hat x}
\newcommand{\hu}{\hat u}
\newcommand{\hv}{\hat v}
\newcommand{\hL}{\hat L}
\newcommand{\hcut}{h^{\mathrm{cut}}}
\newcommand{\hlac}{\hh^{\mathrm{cut}}}
\newcommand{\hint}{\hh^{\mathrm{int}}}
\newcommand{\numstr}{\ell}
\newcommand{\vir}{\emph{Vir}}
\newcommand{\hhgs}{\hh^{\mathrm{gs}}}
\newcommand{\PP}{\mathcal{P}}
\newcommand{\HH}{\mathcal{H}}
\newcommand{\sector}[1]{\HH^{#1}}
\newcommand{\HHtab}{\HH^t_{\ha,\hb}}
\newcommand{\GH}{Y}
\newcommand{\qbinom}[2]{{\genfrac{[}{]}{0pt}{}{#1}{#2}}_q}

\def\CC{{\mathbb{C}}}
\def\ZZ{{\mathbb{Z}}}
\def\HZZ{{\tfrac12\mathbb{Z}}}
\def\HZZp{{\tfrac12\mathbb{Z}_{\ge0}}}

\begin{document}
\title[Half-lattice paths and Virasoro characters]{Half-lattice
              paths and Virasoro characters}
%\thanks{\today}
%\subjclass[2000]{
%        Primary 82B23; Secondary 05A15, 05A19, 17B68, 81T40.}
%\thanks{Research supported by the Research Council (RC)}
%\dedicatory{Version: \today.}

%\author{Olivier Blondeau-Fournier}
\author{Olivier B.-Fournier}
\address{D\'epartement de physique, de g\'enie physique et
d'optique, Universit\'e Laval,  Qu\'ebec, Canada, G1K 7P4.}
\email{olivier.b-fournier.1@ulaval.ca}

\author{Pierre Mathieu}
\address{D\'epartement de physique, de g\'enie physique et
d'optique, Universit\'e Laval,  Qu\'ebec, Canada, G1K 7P4.}
\email{pmathieu@phy.ulaval.ca}

\author{Trevor A Welsh}
\address{
D\'epartement de physique, de g\'enie physique et
d'optique, Universit\'e Laval,  Qu\'ebec, Canada, G1K 7P4;
and
Department of Physics, University of Toronto,
Ontario, Canada, M5S 1A7.}
\email{trevor.welsh@utoronto.ca}

\begin{abstract}
We first briefly review the role of lattice paths in the derivation of
%physically-significant
fermionic expressions for the
$\M(p,p')$ minimal model characters of the Virasoro Lie algebra.
%minimal models that describe universality
%classes of critical points in two dimensions
We then focus on the recently introduced half-lattice paths
for the $\M(p,2p\pm1)$ characters, reformulating them in such a
way that the two cases may be treated uniformly.
That the generating functions of these half-lattice paths are
indeed $\M(p,2p\pm1)$ characters is proved by describing weight
preserving bijections between them and the corresponding RSOS
lattice paths.
Here, the $\M(p,2p-1)$ case is derived for the first time.
We then apply the methods of Bressoud and Warnaar to these
half-lattice paths to derive fermionic expressions for
the Virasoro characters $\chi^{p,2p\pm1}_{1,2}$
that differ from those obtained from the RSOS paths.

This work is an extension of that presented by the third author
at the
``7th International Conference on Lattice Path Combinatorics
     and Applications'', Siena, Italy, July 2010.
\end{abstract}

\maketitle

%\vskip-4mm

%\centerline{\scshape Olivier B.-Fournier%
%\footnote{\label{firstfn}%
%D\'epartement de physique, de g\'enie physique et
%d'optique, Universit\'e Laval,  Qu\'ebec, Canada, G1K 7P4.},
%Pierre Mathieu$^{\ref{firstfn}}$ and
%Trevor A Welsh$^{\ref{firstfn},}$%
%\footnote{
%Department of Physics, University of Toronto,
%Ontario, Canada, M5S 1A7 (Speaker).}%
%}

%\vskip6mm

\section{Introduction}

In this work, we are concerned with the development and exploitation
of combinatorial models for certain $q$-series of importance in
mathematical-physics.
Specifically, we describe various sets of weighted lattice paths,
the generating functions of which are the minimal model Virasoro
characters $\chi^{p,p'}_{r,s}$, where
$p$ and $p'$ are coprime with $1<p<p'$, $1\le r<p$ and $1\le s<p'$.
Explicit expressions for these characters are given by
\begin{equation}
\label{Eq:Rocha}
\chi^{p, p'}_{r, s}=
{\frac{1}{(q)_\infty}}\sum_{\lambda=-\infty}^\infty
(q^{\lambda^2pp'+\lambda(p'r-ps)}-q^{(\lambda p+r)(\lambda p'+s)}),
\end{equation}
where $(q)_\infty=\prod_{i=1}^\infty(1-q^i)$.

One motivation for developing such lattice path models is that they
may be combinatorially manipulated to produce physically significant
\emph{fermionic} expressions \cite{KKMMferm} for the characters
$\chi^{p,p'}_{r,s}$.
These expressions may be interpreted as a sum over the energies of
the excited states of a system of particles for which
%{in particular,}
identical particles are forbidden to occupy the same states.
The following are examples of such expressions:
\begin{subequations}\label{Eq:FermionicEgs}
\begin{align}
\chi^{2,5}_{1,2}&=\sum_{n=0}^\infty \frac{q^{n^2}}{(q)_n},&
\chi^{3,7}_{1,2}&=\sum_{n_1=0}^\infty
\sum_{n_2=0}^\infty \frac{q^{(n_1+n_2)^2+2n_2^2}}{(q)_{n_1}(q)_{2n_2}},\\
%\chi^{3,4}_{1,1}&=\sum_{n=0}^\infty \frac{q^{2n^2}}{(q)_{2n}},&
%\chi^{3,4}_{1,2}&=\sum_{n=0}^\infty \frac{q^{2n^2+n}}{(q)_{2n}}
%                 =\sum_{n=0}^\infty \frac{q^{2n^2-n}}{(q)_{2n+1}},&
\chi^{3,4}_{1,3}&=\sum_{n=0}^\infty \frac{q^{2n^2+2n}}{(q)_{2n+1}},&
\chi^{4,7}_{1,2}&=\sum_{n_1=0}^\infty \sum_{n_2=0}^\infty
\frac{q^{(n_1+2n_2)^2+2n_2^2}}{(q)_{2n_1+4n_2}}
\qbinom{n_1+2n_2}{n_1},
\end{align}
\end{subequations}
where $(q)_n=\prod_{i=1}^n (1-q^i)$ with $(q)_0=1$,
and $\qbinom{m}{n}$ is the $q$-binomial defined by
$\qbinom{m}{n}=(q)_m(q)_n^{-1}(q)_{m-n}^{-1}$ for $0\le n\le m$
and $0$ otherwise.
Such expressions are also of mathematical interest
because they provide the ``sum sides'' of the Rogers-Ramanujan
identities (see \cite{And}, for example) and many generalisations.
Such identities arise in those cases where a product expression
is available for $\chi^{p,p'}_{r,s}$.
For example, three of the cases \eqref{Eq:FermionicEgs} have
product expressions:
\begin{subequations}\label{Eq:ProdEgs}
\begin{align}
\chi^{2,5}_{1,2}&=
\prod^\infty_{\substack{ k=1\\
                         k\equiv \pm 1\bmod5}}
\frac1{1-q^k},
&
\chi^{3,7}_{1,2}&=
\prod^\infty_{\substack{ k=1\\
                         k\not\equiv 0,\pm 2,\pm 10,\pm12,14\bmod28}}
\frac1{1-q^k}
,\\
\chi^{3,4}_{1,3}&=
\prod^\infty_{\substack{ k=1\\
                         k\equiv\pm1,\pm4,\pm6,\pm7\bmod16}}
\frac1{1-q^k}.
\end{align}
\end{subequations}
See \cite{TWfe} for details and references.

The observation \cite{DJKMO} that the generating functions of certain
weighted lattice paths are the minimal model characters
%$\chi^{p,p'}_{r,s}$
of the Virasoro algebra, emerged from the
study of the restricted-solid-on-solid (RSOS) statistical models by
Forrester and Baxter \cite{FB}.
This extended their earlier work with Andrews \cite{ABF},
which dealt with the unitary (ABF) cases, where $p'=p+1$.
\begin{figure}[ht]
\caption{An RSOS path $h\in \PP^{4,9}_{8,6}$.}
\vskip-0.2cm
\label{TypicalScoringFig}
\begin{center}
\psset{unit=0.4cm}
\begin{pspicture}(0,-0.5)(26,9)
%dark bands
\psset{linestyle=none,fillstyle=solid,fillcolor=lightgray}
\psframe(0,2)(26,3)
\psframe(0,4)(26,5)
\psframe(0,6)(26,7)
%dashed grid
\psset{linewidth=0.25pt,linestyle=dashed, dash=2.5pt 1.5pt,linecolor=gray}
\psline{-}(0,2)(26,2) \psline{-}(0,3)(26,3) \psline{-}(0,4)(26,4)
\psline{-}(0,5)(26,5) \psline{-}(0,6)(26,6) \psline{-}(0,7)(26,7)
\psline{-}(1,1)(1,8) \psline{-}(2,1)(2,8) \psline{-}(3,1)(3,8)
\psline{-}(4,1)(4,8) \psline{-}(5,1)(5,8) \psline{-}(6,1)(6,8)
\psline{-}(7,1)(7,8) \psline{-}(8,1)(8,8) \psline{-}(9,1)(9,8)
\psline{-}(10,1)(10,8) \psline{-}(11,1)(11,8) \psline{-}(12,1)(12,8)
\psline{-}(13,1)(13,8) \psline{-}(14,1)(14,8) \psline{-}(15,1)(15,8)
\psline{-}(16,1)(16,8) \psline{-}(17,1)(17,8) \psline{-}(18,1)(18,8)
\psline{-}(19,1)(19,8) \psline{-}(20,1)(20,8) \psline{-}(21,1)(21,8)
\psline{-}(22,1)(22,8) \psline{-}(23,1)(23,8) \psline{-}(24,1)(24,8)
\psline{-}(25,1)(25,8) \psline{-}(26,1)(26,8)
%axes
\psset{linewidth=0.25pt,fillstyle=none,linestyle=solid,linecolor=black}
\psline{->}(0,1)(26.5,1)
\psline{-}(0,8)(26,8)
\psline{-}(0,1)(0,8)
%numeros
\rput(-0.5,1){\scriptsize $1$}\rput(-0.5,2){\scriptsize $2$}
\rput(-0.5,3){\scriptsize $3$}\rput(-0.5,4){\scriptsize $4$}
\rput(-0.5,5){\scriptsize $5$}\rput(-0.5,6){\scriptsize $6$}
\rput(-0.5,7){\scriptsize $7$}\rput(-0.5,8){\scriptsize $8$}
\rput(0,0.5){\scriptsize $0$} \rput(1,0.5){\scriptsize $1$}
\rput(2,0.5){\scriptsize $2$} \rput(3,0.5){\scriptsize $3$}
\rput(4,0.5){\scriptsize $4$} \rput(5,0.5){\scriptsize $5$}
\rput(6,0.5){\scriptsize $6$} \rput(7,0.5){\scriptsize $7$}
\rput(8,0.5){\scriptsize $8$} \rput(9,0.5){\scriptsize $9$}
\rput(10,0.5){\scriptsize $10$} \rput(11,0.5){\scriptsize $11$}
\rput(12,0.5){\scriptsize $12$} \rput(13,0.5){\scriptsize $13$}
\rput(14,0.5){\scriptsize $14$} \rput(15,0.5){\scriptsize $15$}
\rput(16,0.5){\scriptsize $16$} \rput(17,0.5){\scriptsize $17$}
\rput(18,0.5){\scriptsize $18$} \rput(19,0.5){\scriptsize $19$}
\rput(20,0.5){\scriptsize $20$} \rput(21,0.5){\scriptsize $21$}
\rput(22,0.5){\scriptsize $22$} \rput(23,0.5){\scriptsize $23$}
\rput(24,0.5){\scriptsize $24$} \rput(25,0.5){\scriptsize $25$}
\rput(26,0.5){\scriptsize $26$}
%path
\psset{linewidth=0.7pt,fillstyle=none,linestyle=solid,linecolor=black}
\psline(0,8)(1,7)(2,6)(3,5)(4,6)(5,5)(6,4)(7,3)(8,2)(9,3)(10,2)(11,1)
 (12,2)(13,3)(14,4)(15,5)(16,4)(17,3)(18,4)(19,5)(20,6)(21,5)(22,6)
 (23,7)(24,6)(25,7)(26,6)
%dots
\psset{dotsize=3.5pt,fillcolor=black}
\psdots(1,7)(3,5)(5,5)(7,3)(11,1)(17,3)(21,5)
\psset{dotsize=4pt,dotstyle=o,fillcolor=white}
\psdots(4,6)(12,2)(14,4)(18,4)(20,6)(22,6)
\end{pspicture}
\end{center}
\end{figure}
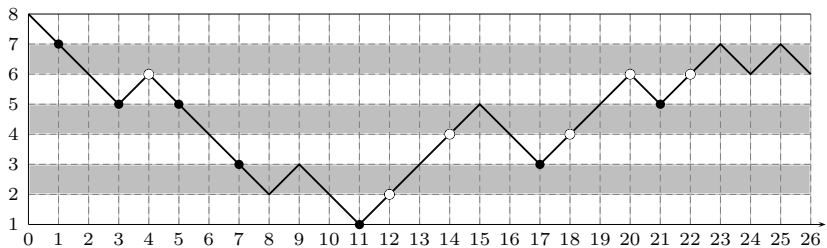
Below, in Section \ref{Sec:RSOS}, we review a reformulation
of these paths due to Foda \emph{et al.} \cite{FLPW}, which has
proved extremely useful in the derivation of fermionic expressions.
A typical RSOS path is shown in Fig.~\ref{TypicalScoringFig}.

We then describe alternative sets of lattice paths,
which have been recently introduced \cite{PMnpd,BfMW},
whose generating functions turn out to be Virasoro characters
in the non-unitary cases for which $p'=2p\pm1$
(the $p'=2p-1$ case was sketched in the concluding section of \cite{BfMW}).
Significantly, these lattice paths,
which we refer to as \emph{half-lattice} paths,
share some of the simplifying characteristics of the
ABF paths, permitting easier analysis.
%underlying statistical model interpretation?
%(the 1991 paper by Kuniba (NPB355, 801--821) might give it)
That the generating functions of these half-lattice paths
are Virasoro characters may be proved by formulating
weight-preserving bijections between them and the corresponding
RSOS paths, the generating functions for which
are known to be the $\chi^{p,p'}_{r,s}$ \cite{FLPW}.
We describe such bijections in Sections \ref{Sec:Bijection1}
and \ref{Sec:Bijection2} below.

We then show how these half-lattice paths may be manipulated using
the techniques of Bressoud \cite{Bres} and Warnaar \cite{OleJS1}
to yield fermionic expressions for their generating functions.
Thereupon, through the above bijection, we have obtained
fermionic expressions for Virasoro characters
$\chi^{p,p'}_{r,s}$
in the cases for which $p'=2p\pm1$, $r=1$ and $s=2$.
These expressions were previously conjectured by Berkovich, McCoy and
Pearce \cite[\S9]{BMP}, and proved by Warnaar \cite[\S IVD]{OleJMP99}.
We will consider other cases of $r$ and $s$ elsewhere.

A brief review of the Virasoro algebra and its $\M(p,p')$
minimal model representations is given in an appendix.
More details may be found in \cite{KR,DfMS}.

\section{RSOS paths}
\label{Sec:RSOS}

\subsection{Specification}

Here, we define an RSOS path to be an infinite integer sequence
$h=(h_0,h_1,h_2,\ldots)$ for which $|h_{x+1}-h_x|=1$ for each $x\ge0$.
An RSOS path $h$ is said to be $(f,g)$-restricted if
$f\le h_x\le g$ for all $x\ge0$, and $b$-tailed if there exists
$L$ such that $h_x\in\{b,b+1\}$ for $x\ge L$.
For $p$ and $p'$ coprime with $1<p<p'$, and $1\le a,b<p'$,
define $\PP^{p,p'}_{a,b}$ to be the set of those RSOS paths $h$
which are $(1,p'-1)$-restricted, $b$-tailed, and have $h_0=a$.

The \emph{path picture} of an RSOS path
$h\in\PP^{p,p'}_{a,b}$ is obtained by linking the vertices
$(0,h_0)$, $(1,h_1)$, $(2,h_2),\ldots$ on the plane;
it has an infinite tail oscillating between adjacent heights $b$ and $b+1$.
%An example of an RSOS path is given in Fig.~\ref{TypicalScoringFig}
%(the significance of the shading, circles and dots is explained below).
An example for an RSOS path, its tail truncated,
is given in Fig.~\ref{TypicalScoringFig}
(the significance of the shading, circles and dots is explained below).

A vertex $(x,h_x)$ is said to be straight if, in the path picture,
its two neighbouring edges are in the same direction,
and a peak or a valley if those neighbouring edges
are in the NE-SE or SE-NE directions respectively.

For $h\in\PP^{p,p'}_{a,b}$, it is convenient,
for the present purposes, to define $L(h)$ to be
the smallest value of $L\in 2\ZZ$ for which $h_x\in\{b,b+1\}$
for all $x\ge L$.
%We refer to $L(h)$ as the effective length of $h$.
% Then $h_{L(h)}=b$.  % Not true.

\subsection{Weighting RSOS paths}
\label{Sec:RSOSWts}

When analysing the RSOS paths from $\PP^{p,p'}_{a,b}$,
it is convenient to shade the horizontal band of the plane
between adjacent heights
$\lfloor rp'/p\rfloor$ and $\lfloor rp'/p\rfloor+1$ for
each $r$ with $1\le r<p$.
Such bands are referred to as dark bands, with the others being
light bands.

Given a path $h\in\PP^{p,p'}_{a,b}$, a vertex $(x,h_x)$ is
said to be \emph{scoring} if either it is straight
with the right edge in a dark band, or it is not straight
with the right edge in a light band,
All other vertices are \emph{non-scoring} vertices.
Each scoring vertex is designated \emph{up-scoring} or
\emph{down-scoring} depending on whether the direction of
the left edge is NE or SE respectively.
In the path picture, we highlight each up-scoring vertex
with an unfilled circle, and each down-scoring vertex with a
filled circle.
For example, see Fig.~\ref{TypicalScoringFig}.

For each vertex $(x,h_x)$ we set
\begin{equation}\label{Eq:ScoreWts}
u_x=\frac12 (x-h_x+a),\qquad
v_x=\frac12 (x+h_x-a).
\end{equation}
The weight $\wt(h)$ of $h\in\PP^{p,p'}_{a,b}$ is then defined to
be the sum over the up-scoring vertices of their values of $u_x$,
plus the sum over the down-scoring vertices of their values of $v_x$:
\begin{equation}
\label{Eq:RSOSWtDef}
\wt(h)
=\sum_{\begin{subarray}{c} \text{up-scoring}\\
                           (x,h_x)
       \end{subarray}} \!\! u_x
+\sum_{\begin{subarray}{c} \text{down-scoring}\\
                           (x,h_x)
       \end{subarray}} \!\!\!\!  v_x.
\end{equation}
For example, on enumerating the scoring vertices of the path
$h$ in Fig.~\ref{TypicalScoringFig} from left to right, we
obtain $\wt(h)=0+0+3+1+1+2+9+9+6+11+11+9+12=74$.
This definition of $\wt(h)$ is due to
Foda \emph{et al.} \cite{FLPW}, and is a considerable simplification
on that originally given in \cite{FB}.
Note that $\wt(h)$ is finite only if $b=\lfloor rp'/p\rfloor$
for $r$ satisfying $1\le r<p$, for otherwise the infinite tail
of $h$ lies in a light band, whereupon each of its vertices
has a positive contribution to $\wt(h)$.

The generating function $X^{p,p'}_{a,b}(q)$ for the set
$\PP^{p,p'}_{a,b}$ of RSOS paths is then defined by
\begin{equation}\label{Eq:XGFdef}
X^{p,p'}_{a,b}(q)=\sum_{h\in\PP^{p,p'}_{a,b}} q^{\wt(h)}.
\end{equation}

It may be shown that if $b=\lfloor rp'/p\rfloor$ then
\begin{equation}\label{Eq:XRocha}
X^{p, p'}_{a, b}(q)=
\chi^{p,p'}_{r,a},
\end{equation}
where the Virasoro character $\chi^{p,p'}_{r,a}$ is given
explicitly in \eqref{Eq:Rocha}.
This result was proved in \cite{FLPW} using recurrence relations
for the paths.
It may also be proved using an inclusion-exclusion argument.
Because of this result, the RSOS lattice paths may be used as
convenient labels for the states of the minimal model Virasoro modules.

\subsection{Obtaining Fermionic expressions}

In those cases for which $p'=p+1$ (the ABF cases),
the weighting rule degenerates to that originally given in \cite{ABF},
where $\wt(h)$ is obtained simply by summing half the
$x$ positions of the straight vertices of the path $h$.
This weighting rule was necessary for Warnaar \cite{OleJS1,OleJS2}
to give a direct segmentation of a path,
from which fermionic expressions for each $\chi^{p,p+1}_{r,s}$
could be immediately deduced.

Fermionic expressions are not so readily derived in the
general $p'>p+1$ cases.
However, a combinatorial means of obtaining a particle description
recursively was devised by Foda \emph{et al.} \cite{FLPW},
leading to fermionic
expressions for $\chi^{p,p'}_{r,s}$ in all cases in which
$s$ and $r$ are respectively
a Takahashi length and a truncated Takahashi length \cite{FW-kyoto}
(these expressions were previously obtained in \cite{BM,BMS}).
These methods were subsequently extended to yield fermionic
expressions for all $\chi^{p,p'}_{r,s}$ \cite{TWfe}.
An alternative, non-recursive means of obtaining the same
particle description in the $p'\ge2p-1$ cases was outlined
in \cite{JMrsos}.

\section{Half-lattice paths}

\subsection{Introduction}

Recently, new lattice path models for the minimal models
\hbox{$\M(p,2p+1)$} and $\M(p,2p-1)$ have been given \cite{PMnpd,BfMW}.
The description that follows recasts those descriptions so
that the two cases can be treated in a uniform way.
To prove that the generating functions in these two cases are the
Virasoro characters, we formulate bijections between the half-lattice
paths and the corresponding RSOS paths for these minimal models in
Sections \ref{Sec:Bijection1} and \ref{Sec:Bijection2} respectively.
The first of these bijections is a recasting of that described
in \cite{BfMW}.
The other is new.

\subsection{Specification}
\label{Sec:Spec}

A half-lattice path is defined to be
an infinite sequence $\hh=(\hh_0,\hh_{1/2},\hh_1,$ $\hh_{3/2},\ldots)$
satisfying $\hh_x\in\HZZ$ and
$|\hh_{x+1/2}-\hh_x|=\tfrac12$ for each $x\in\HZZp$,
with the \emph{additional} restriction that if
$\hh_x=\hh_{x+1}\in\ZZ$, then $\hh_{x+1/2}=\hh_x+1/2$.
A half-lattice path $\hh$ is said to be $(f,g)$-restricted if
$f\le \hh_x\le g$ for all $x\ge0$, and $\hb$-tailed if there exists
$L$ such that $\hh_x\in\{\hb,\hb+\tfrac12\}$ for $x\ge L$.
%Similar to the RSOS case,
Here, we define sets of half-lattice paths for each $t\in\HZZ$
with $t\ge2$.
For $\ha,\hb\in\ZZ$, define $\HHtab$ to be the
set of all half-lattice paths $\hh$ that are $(1,t)$-restricted,
$\hb$-tailed, with $\hh_0=\ha$.

The \emph{path picture} of a restricted half-lattice path
$\hh\in\HHtab$ is obtained by linking the vertices
$(0,\hh_0), (1/2,\hh_{1/2}), (1,\hh_1),\ldots$ on the plane;
it has an infinite oscillating tail.
Examples for $t=4$ and $t=7/2$, their tails truncated, are given in
Figs.~\ref{TypicalHalfHat} and \ref{TypicalDualBij} respectively.

A vertex $(x,\hh_x)$ is said to be a peak, a valley,
straight-up or straight-down, depending on whether the pair
of edges that neighbour $(x,\hh_x)$ in this path picture are
in the NE-SE, SE-NE, NE-NE, or SE-SE directions respectively.
In order to likewise specify the nature of the vertex $(0,\ha)$
at the path's startpoint, we adopt the convention that
$\hh_{-1/2}=\ha+1/2$.
%(doing so even if $\ha=t$).
Note that the additional restriction above implies that valleys can
only occur at integer heights.

For $\hh\in\HHtab$, we define $\hL(\hh)$ to be the smallest value
of $L\in \ZZ$ for which $\hh_x\in\{\hb,\hb+\tfrac12\}$
for all $x\ge L$. Then $\hh_{\hL(\hh)}=\hb$.

The \emph{unnormalised} weight $\unwt(\hh)$ of a half-lattice
path $\hh$ is defined to be half the sum of those
$x\in\tfrac12{\ZZ_{\ge0}}$ for which $(x,\hh_x)$ is a straight-vertex:
\begin{equation}
\label{Eq:WtsDef}
%\unwt(\hh)=\frac12 \sum_{i=1}^{\infty} |h_{(i+1)/2}-h_{(i-1)/2}|.
\unwt(\hh)
=\frac12\sum_{\begin{subarray}{c} x\in\frac12{\ZZ_{\ge0}}\\
                                  \hh_{x-1/2}\ne \hh_{x+1/2}
              \end{subarray}} \!\! x.
\end{equation}
The \emph{ground-state path}
$\hhgs\in\HHtab$ is defined to be that path which has minimal weight
amongst all the elements of $\HHtab$
(this path is simply a straight line linking height $\ha$ to $\hb$,
followed by a $\hb$-tail).
The weight $\wh(\hh)$ of $\hh$ is then defined by
\begin{equation}\label{Eq:HalfWt}
\wh(\hh)=\unwt(\hh)-\unwt(\hhgs).
\end{equation}

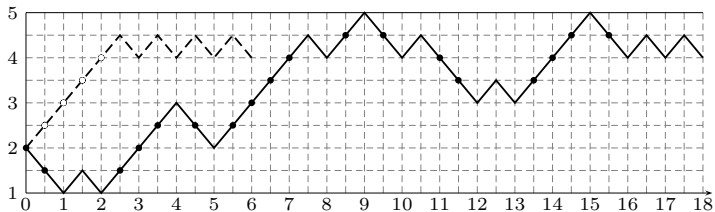
\begin{figure}[ht]
\caption{Half-lattice path $\hh\in\HH^{5}_{2,4}$ (solid)
          and $\hhgs\in\HH^{5}_{2,4}$ (dashed)}
\vskip-1mm
\label{Fig:SolidNDashed}
\begin{center}
\psset{yunit=0.30cm,xunit=0.25cm}
\begin{pspicture}(0,-0.5)(36,10)
%dashed grid
\psset{linewidth=0.25pt,linestyle=dashed, dash=2.5pt 1.5pt,linecolor=gray}
\psline{-}(0,2)(36,2) \psline{-}(0,3)(36,3) \psline{-}(0,4)(36,4)
\psline{-}(0,5)(36,5) \psline{-}(0,6)(36,6) \psline{-}(0,7)(36,7)
\psline{-}(0,8)(36,8)
\psline{-}(1,1)(1,9) \psline{-}(2,1)(2,9) \psline{-}(3,1)(3,9)
\psline{-}(4,1)(4,9) \psline{-}(5,1)(5,9) \psline{-}(6,1)(6,9)
\psline{-}(7,1)(7,9) \psline{-}(8,1)(8,9) \psline{-}(9,1)(9,9)
\psline{-}(10,1)(10,9) \psline{-}(11,1)(11,9) \psline{-}(12,1)(12,9)
\psline{-}(13,1)(13,9) \psline{-}(14,1)(14,9) \psline{-}(15,1)(15,9)
\psline{-}(16,1)(16,9) \psline{-}(17,1)(17,9) \psline{-}(18,1)(18,9)
\psline{-}(19,1)(19,9) \psline{-}(20,1)(20,9) \psline{-}(21,1)(21,9)
\psline{-}(22,1)(22,9) \psline{-}(23,1)(23,9) \psline{-}(24,1)(24,9)
\psline{-}(25,1)(25,9) \psline{-}(26,1)(26,9) \psline{-}(27,1)(27,9)
\psline{-}(28,1)(28,9) \psline{-}(29,1)(29,9) \psline{-}(30,1)(30,9)
\psline{-}(31,1)(31,9) \psline{-}(32,1)(32,9) \psline{-}(33,1)(33,9)
\psline{-}(34,1)(34,9) \psline{-}(35,1)(35,9) \psline{-}(36,1)(36,9)
%axes
\psset{linewidth=0.25pt,fillstyle=none,linestyle=solid,linecolor=black}
\psline{->}(0,1)(36.5,1)
\psline{-}(0,9)(36,9)
\psline{-}(0,1)(0,9)
%numeros
\rput(-0.7,1){\scriptsize $1$}\rput(-0.7,3){\scriptsize $2$}
\rput(-0.7,5){\scriptsize $3$} \rput(-0.7,7){\scriptsize $4$}
\rput(-0.7,9){\scriptsize $5$}
\rput(0,0.5){\scriptsize $0$} \rput(2,0.5){\scriptsize $1$}
\rput(4,0.5){\scriptsize $2$} \rput(6,0.5){\scriptsize $3$}
\rput(8,0.5){\scriptsize $4$} \rput(10,0.5){\scriptsize $5$}
\rput(12,0.5){\scriptsize $6$} \rput(14,0.5){\scriptsize $7$}
\rput(16,0.5){\scriptsize $8$} \rput(18,0.5){\scriptsize $9$}
\rput(20,0.5){\scriptsize $10$} \rput(22,0.5){\scriptsize $11$}
\rput(24,0.5){\scriptsize $12$} \rput(26,0.5){\scriptsize $13$}
\rput(28,0.5){\scriptsize $14$} \rput(30,0.5){\scriptsize $15$}
\rput(32,0.5){\scriptsize $16$} \rput(34,0.5){\scriptsize $17$}
\rput(36,0.5){\scriptsize $18$}
%path
\psset{linewidth=0.7pt,fillstyle=none,linestyle=solid,linecolor=black}
\psline(0,3)(1,2)(2,1)(3,2)(4,1)(5,2)(6,3)(7,4)(8,5)(9,4)(10,3)(11,4)
 (12,5)(13,6)(14,7)(15,8)(16,7)(17,8)(18,9)(19,8)(20,7)(21,8)(22,7)
 (23,6)(24,5)(25,6)(26,5)(27,6)(28,7)(29,8)(30,9)(31,8)(32,7)(33,8)
 (34,7)(35,8)(36,7)
%dots
\psset{dotsize=2.5pt,fillcolor=black}
\psdots(0,3)(1,2)(5,2)(6,3)(7,4)(9,4)(11,4)(12,5)(13,6)(14,7)
(17,8)(19,8)(22,7)(23,6)(27,6)(28,7)(29,8)(31,8)
%gs
\psset{linewidth=0.7pt,linestyle=dashed, dash=4pt 2pt,linecolor=black}
\psline(0,3)(1,4)(2,5)(3,6)(4,7)(5,8)(6,7)(7,8)(8,7)(9,8)(10,7)(11,8)(12,7)
\psset{dotsize=2.5pt,dotstyle=o,fillcolor=white}
\psdots(1,4)(2,5)(3,6)(4,7)
\end{pspicture}
\end{center}
\end{figure}

\noindent
To illustrate this definition, consider the path $\hh\in\HH^{5}_{2,4}$
which is depicted as the solid line in Fig.~\ref{Fig:SolidNDashed}.
Its straight vertices are highlighted here using dots.
These are the vertices that contribute to $\unwt(\hh)$ in
\eqref{Eq:WtsDef}, giving
\begin{equation}
\begin{split}
\unwt(\hh)&=\tfrac12\big(
0+
\tfrac12+\tfrac52+3+\tfrac72+\tfrac92+\tfrac{11}2+6+\tfrac{13}2+7
+\tfrac{17}2\\
&\hskip35mm
+\tfrac{19}2+11+\tfrac{23}2+\tfrac{27}2+14+\tfrac{29}2+\tfrac{31}2
\big)
=\tfrac{137}2.
\end{split}
\end{equation}
The corresponding ground-state path $\hhgs\in\HH^{5}_{2,4}$ is shown
as the dashed line in Fig.~\ref{Fig:SolidNDashed}.
Its straight vertices are highlighted here using circles.
These lead to
\begin{equation}
\unwt(\hhgs)=\tfrac12\big(\tfrac12+1+\tfrac32+2\big)=\tfrac52.
\end{equation}
Thereupon \eqref{Eq:HalfWt} yields $\wh(\hh)=\tfrac12(137-5)=66$.

We define the generating functions for the sets $\HHtab$ of half-lattice
paths by
\begin{equation}\label{Eq:YGFdef}
\GH^t_{\ha,\hb}(q)=
\sum_{\hh\in\HHtab} q^{\wh(\hh)}.
\end{equation}
Our first main result is that these generating functions are Virasoro
characters:

\begin{theorem}\label{Thm:HalfChar}
If $t\in\ZZ_{\ge2}$, and $\hat a\in\{1,2,\ldots,t\}$ and
$\hat b\in\{1,2,\ldots,t-1\}$, then
\begin{equation}
\label{Eq:HalfChar1}
\GH^t_{\hat a,\hat b}(q)=\chi^{t,2t+1}_{\hat b,2\hat a}.
\end{equation}
If $t\in\ZZ_{\ge2}+\frac12$, and
$\hat a,\hat b\in\{1,2,\ldots,t-\frac12\}$ then
\begin{equation}
\label{Eq:HalfChar2}
\GH^t_{\hat a,\hat b}(q)=\chi^{t+1/2,2t}_{\hat a,2\hat b}.
\end{equation}
\end{theorem}
Note that by using the identity \eqref{Eq:RCid1}
we can see that each $\M(p,2p+1)$ character is one
$\GH^p_{\hat a,\hat b}(q)$,
and each $\M(p,2p-1)$ character is one
$\GH^{p-1/2}_{\hat a,\hat b}(q)$.
Also note that identity \eqref{Eq:RCid2} enables
Theorem \ref{Thm:HalfChar} to be simplified by
extending \eqref{Eq:HalfChar1} to apply to all $t\in\frac12\ZZ_{\ge4}$
since for the non-integer cases of $t$, \eqref{Eq:HalfChar2} is then
subsumed into \eqref{Eq:HalfChar1}.

In what follows, Theorem \ref{Thm:HalfChar} is proved by establishing
weight-preserving bijections between the sets of half-lattice paths
and the corresponding sets of RSOS paths.

\subsection{Useful trick}
\label{Sec:WtTrick}

Before we proceed, we describe a useful trick to obtain
$\wh(\hh)$ directly from the path picture of
$\hh\in\HHtab$.
First note that the minimal weight path
$\hhgs\in\HHtab$
extends between heights $\ha$ and $\hb$
in its first $2e$ (half-integer) steps, where $e=|\ha-\hb|$.
An alternative to \eqref{Eq:HalfWt} for obtaining $\wh(\hh)$,
is to extend $\hh$
to the left by $2e$ steps, in such a way that $\hh_{-e}=\hb$
(overriding the above convention for $\hh_{-1/2}$).
The renormalised weight $\wh(\hh)$ is then obtained by summing the
$x$-coordinates of all the straight vertices of this extended path,
beginning with its first vertex at $(-e,\hat b)$,
whose nature is specified by setting $\hh_{-e-1/2}=\hb+1/2$,
and dividing by 2.

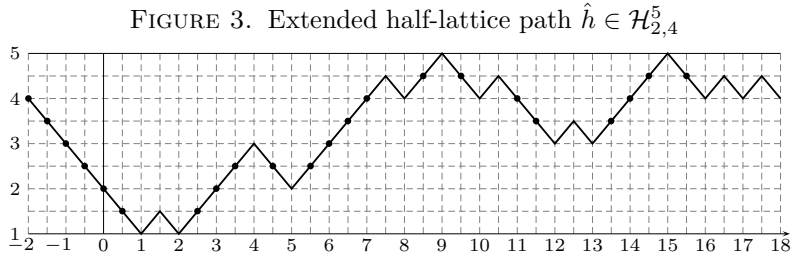
\begin{figure}[ht]
\caption{Extended half-lattice path $\hh\in\HH^{5}_{2,4}$}
\vskip-1mm
\label{Fig:Extended}
\begin{center}
\psset{yunit=0.30cm,xunit=0.25cm}
\begin{pspicture}(-4,-0.5)(36,10)
%dashed grid
\psset{linewidth=0.25pt,linestyle=dashed, dash=2.5pt 1.5pt,linecolor=gray}
\psline{-}(-4,2)(36,2) \psline{-}(-4,3)(36,3) \psline{-}(-4,4)(36,4)
\psline{-}(-4,5)(36,5) \psline{-}(-4,6)(36,6) \psline{-}(-4,7)(36,7)
\psline{-}(-4,8)(36,8)
\psline{-}(-4,1)(-4,9) \psline{-}(-3,1)(-3,9) \psline{-}(-2,1)(-2,9)
\psline{-}(-1,1)(-1,9)
\psline{-}(1,1)(1,9) \psline{-}(2,1)(2,9) \psline{-}(3,1)(3,9)
\psline{-}(4,1)(4,9) \psline{-}(5,1)(5,9) \psline{-}(6,1)(6,9)
\psline{-}(7,1)(7,9) \psline{-}(8,1)(8,9) \psline{-}(9,1)(9,9)
\psline{-}(10,1)(10,9) \psline{-}(11,1)(11,9) \psline{-}(12,1)(12,9)
\psline{-}(13,1)(13,9) \psline{-}(14,1)(14,9) \psline{-}(15,1)(15,9)
\psline{-}(16,1)(16,9) \psline{-}(17,1)(17,9) \psline{-}(18,1)(18,9)
\psline{-}(19,1)(19,9) \psline{-}(20,1)(20,9) \psline{-}(21,1)(21,9)
\psline{-}(22,1)(22,9) \psline{-}(23,1)(23,9) \psline{-}(24,1)(24,9)
\psline{-}(25,1)(25,9) \psline{-}(26,1)(26,9) \psline{-}(27,1)(27,9)
\psline{-}(28,1)(28,9) \psline{-}(29,1)(29,9) \psline{-}(30,1)(30,9)
\psline{-}(31,1)(31,9) \psline{-}(32,1)(32,9) \psline{-}(33,1)(33,9)
\psline{-}(34,1)(34,9) \psline{-}(35,1)(35,9) \psline{-}(36,1)(36,9)
%axes
\psset{linewidth=0.25pt,fillstyle=none,linestyle=solid,linecolor=black}
\psline{->}(-4,1)(36.5,1)
\psline{-}(-4,9)(36,9)
\psline{-}(0,1)(0,9)
%numeros
\rput(-4.7,1){\scriptsize $1$}\rput(-4.7,3){\scriptsize $2$}
\rput(-4.7,5){\scriptsize $3$} \rput(-4.7,7){\scriptsize $4$}
\rput(-4.7,9){\scriptsize $5$}
\rput(-4.4,0.5){\scriptsize $-2$} \rput(-2.4,0.5){\scriptsize $-1$}
\rput(0,0.5){\scriptsize $0$} \rput(2,0.5){\scriptsize $1$}
\rput(4,0.5){\scriptsize $2$} \rput(6,0.5){\scriptsize $3$}
\rput(8,0.5){\scriptsize $4$} \rput(10,0.5){\scriptsize $5$}
\rput(12,0.5){\scriptsize $6$} \rput(14,0.5){\scriptsize $7$}
\rput(16,0.5){\scriptsize $8$} \rput(18,0.5){\scriptsize $9$}
\rput(20,0.5){\scriptsize $10$} \rput(22,0.5){\scriptsize $11$}
\rput(24,0.5){\scriptsize $12$} \rput(26,0.5){\scriptsize $13$}
\rput(28,0.5){\scriptsize $14$} \rput(30,0.5){\scriptsize $15$}
\rput(32,0.5){\scriptsize $16$} \rput(34,0.5){\scriptsize $17$}
\rput(36,0.5){\scriptsize $18$}
%path
\psset{linewidth=0.7pt,fillstyle=none,linestyle=solid,linecolor=black}
\psline(-4,7)(-3,6)(-2,5)(-1,4)
 (0,3)(1,2)(2,1)(3,2)(4,1)(5,2)(6,3)(7,4)(8,5)(9,4)(10,3)(11,4)
 (12,5)(13,6)(14,7)(15,8)(16,7)(17,8)(18,9)(19,8)(20,7)(21,8)(22,7)
 (23,6)(24,5)(25,6)(26,5)(27,6)(28,7)(29,8)(30,9)(31,8)(32,7)(33,8)
 (34,7)(35,8)(36,7)
%dots
\psset{dotsize=2.5pt,fillcolor=black}
\psdots(-4,7)(-3,6)(-2,5)(-1,4)
(0,3)(1,2)(5,2)(6,3)(7,4)(9,4)(11,4)(12,5)(13,6)(14,7)
(17,8)(19,8)(22,7)(23,6)(27,6)(28,7)(29,8)(31,8)
\end{pspicture}
\end{center}
\end{figure}

To illustrate this construction, consider the path
$\hh\in\HH^5_{2,4}$ represented by the solid line in
Fig.~\ref{Fig:SolidNDashed},
for which we found $\wh(\hh)=66$.
In this case, the extended path described above is shown
in Fig.~\ref{Fig:Extended}.
From this, the renormalised weight $\wh(\hh)$ is immediately
obtained from its straight vertices via
\begin{equation}
\begin{split}
\wh(\hh)&=\tfrac12\big(
-2-\tfrac32-1-\tfrac12+0+
\tfrac12+\tfrac52+3+\tfrac72+\tfrac92+\tfrac{11}2+6+\tfrac{13}2\\
&\hskip30mm
+7+\tfrac{17}2
+\tfrac{19}2+11+\tfrac{23}2+\tfrac{27}2+14+\tfrac{29}2+\tfrac{31}2
\big)
=66.
\end{split}
\end{equation}

\section{The $\M(p,2p+1)$ bijection}
\label{Sec:Bijection1}

\subsection{Description}

Here, we describe a weight-preserving bijection
\begin{equation}\label{Eq:Bijection1}
\PP^{p,2p+1}_{a,b}\leftrightarrow \HH^p_{a/2,b/2} ,
\end{equation}
for $p\ge2$, and $a$ and $b$ even
integers with $1<a\le 2p$ and $1<b<2p$.
From this, via \eqref{Eq:XGFdef}, \eqref{Eq:XRocha} and \eqref{Eq:YGFdef},
the identity \eqref{Eq:HalfChar1} is immediately established.

\begin{figure}[ht]
\caption{$h^{\text{cut}}$ obtained from Fig. \ref{TypicalScoringFig}.}
\vskip-0.2cm
\label{TypicalScoringCut}
\begin{center}
\psset{unit=0.4cm}

\begin{pspicture}(0,-0.5)(26,9)
%dark bands
\psset{linestyle=none,fillstyle=solid,fillcolor=lightgray}
\psframe(0,2)(26,3)
\psframe(0,4)(26,5)
\psframe(0,6)(26,7)
%dashed grid
\psset{linewidth=0.25pt,linestyle=dashed, dash=2.5pt 1.5pt,linecolor=gray}
\psline{-}(0,2)(26,2) \psline{-}(0,3)(26,3) \psline{-}(0,4)(26,4)
\psline{-}(0,5)(26,5) \psline{-}(0,6)(26,6) \psline{-}(0,7)(26,7)
\psline{-}(1,1)(1,8) \psline{-}(2,1)(2,8) \psline{-}(3,1)(3,8)
\psline{-}(4,1)(4,8) \psline{-}(5,1)(5,8) \psline{-}(6,1)(6,8)
\psline{-}(7,1)(7,8) \psline{-}(8,1)(8,8) \psline{-}(9,1)(9,8)
\psline{-}(10,1)(10,8) \psline{-}(11,1)(11,8) \psline{-}(12,1)(12,8)
\psline{-}(13,1)(13,8) \psline{-}(14,1)(14,8) \psline{-}(15,1)(15,8)
\psline{-}(16,1)(16,8) \psline{-}(17,1)(17,8) \psline{-}(18,1)(18,8)
\psline{-}(19,1)(19,8) \psline{-}(20,1)(20,8) \psline{-}(21,1)(21,8)
\psline{-}(22,1)(22,8) \psline{-}(23,1)(23,8) \psline{-}(24,1)(24,8)
\psline{-}(25,1)(25,8) \psline{-}(26,1)(26,8)
%axes
\psset{linewidth=0.25pt,fillstyle=none,linestyle=solid,linecolor=black}
\psline{->}(0,1)(26.5,1)
\psline{-}(0,8)(26,8)
\psline{-}(0,1)(0,8)
%numeros
\rput(-0.5,1){\scriptsize $1$}\rput(-0.5,2){\scriptsize $2$}
\rput(-0.5,3){\scriptsize $3$}\rput(-0.5,4){\scriptsize $4$}
\rput(-0.5,5){\scriptsize $5$}\rput(-0.5,6){\scriptsize $6$}
\rput(-0.5,7){\scriptsize $7$}\rput(-0.5,8){\scriptsize $8$}
\rput(0,0.5){\scriptsize $0$} \rput(1,0.5){\scriptsize $1$}
\rput(2,0.5){\scriptsize $2$} \rput(3,0.5){\scriptsize $3$}
\rput(4,0.5){\scriptsize $4$} \rput(5,0.5){\scriptsize $5$}
\rput(6,0.5){\scriptsize $6$} \rput(7,0.5){\scriptsize $7$}
\rput(8,0.5){\scriptsize $8$} \rput(9,0.5){\scriptsize $9$}
\rput(10,0.5){\scriptsize $10$} \rput(11,0.5){\scriptsize $11$}
\rput(12,0.5){\scriptsize $12$} \rput(13,0.5){\scriptsize $13$}
\rput(14,0.5){\scriptsize $14$} \rput(15,0.5){\scriptsize $15$}
\rput(16,0.5){\scriptsize $16$} \rput(17,0.5){\scriptsize $17$}
\rput(18,0.5){\scriptsize $18$} \rput(19,0.5){\scriptsize $19$}
\rput(20,0.5){\scriptsize $20$} \rput(21,0.5){\scriptsize $21$}
\rput(22,0.5){\scriptsize $22$} \rput(23,0.5){\scriptsize $23$}
\rput(24,0.5){\scriptsize $24$} \rput(25,0.5){\scriptsize $25$}
\rput(26,0.5){\scriptsize $26$}
%path
\psset{linewidth=0.7pt,fillstyle=none,linestyle=solid,linecolor=black}
\psline(0,8)(1,7)(2,6)(3,5)(4,4)(5,3)(6,2)(7,3)(8,2)(9,3)(10,4)(11,5)
 (12,4)(13,5)(14,6)(15,7)(16,6)(17,7)(18,6)(19,7)(20,6)(21,7)(22,6)
 (23,7)(24,6)(25,7)(26,6)
%dots
\psset{dotsize=3.5pt,fillcolor=black}
\psdots(1,7)(3,5)(5,3)
\psset{dotsize=4pt,dotstyle=o,fillcolor=white}
\psdots(10,4)(14,6)
\end{pspicture}
\end{center}
\end{figure}
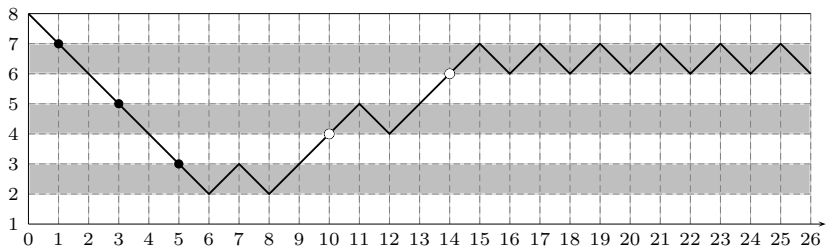

Let $h\in\PP^{p,2p+1}_{a,b}$, and note that in this case, the
RSOS band structure has alternating white and dark bands,
with the uppermost and lowermost bands both white.
Each pair of adjacent scoring vertices is said to be a particle
(for $d$ adjacent scoring vertices, there are $\lfloor d/2\rfloor$
particles).
For the $i$th such particle of $h$, counting from the right,
let $\lambda_i$ be the number of non-scoring vertices to its left.
If $h$ has $n$ particles,
$\lambda=(\lambda_1,\lambda_2,\ldots,\lambda_n)$ is a partition.
Now create the path $\hcut\in\PP^{p,2p+1}_{a,b}$, by removing
all the particles from $h$, in each case joining up the loose
ends, which will be of the same height.
In the case of the path from Fig.~\ref{TypicalScoringFig},
this yields the path $\hcut$ given in Fig.~\ref{TypicalScoringCut},
with $\lambda=(9,8,5,1)$.

\begin{figure}[ht]
\caption{Half-lattice path $\hlac\in\mathcal{H}_{4,3}^{4}$
  obtained from Fig.~\ref{TypicalScoringCut}}
\vskip-1mm
\label{TypicalScoringHat}
\begin{center}
\psset{yunit=0.32cm,xunit=0.28cm}

\begin{pspicture}(0,-0.5)(46,8)
%dashed grid
\psset{linewidth=0.25pt,linestyle=dashed, dash=2.5pt 1.5pt,linecolor=gray}
\psline{-}(0,2)(46,2) \psline{-}(0,3)(46,3) \psline{-}(0,4)(46,4)
\psline{-}(0,5)(46,5) \psline{-}(0,6)(46,6)
\psline{-}(1,1)(1,7) \psline{-}(2,1)(2,7) \psline{-}(3,1)(3,7)
\psline{-}(4,1)(4,7) \psline{-}(5,1)(5,7) \psline{-}(6,1)(6,7)
\psline{-}(7,1)(7,7) \psline{-}(8,1)(8,7) \psline{-}(9,1)(9,7)
\psline{-}(10,1)(10,7) \psline{-}(11,1)(11,7) \psline{-}(12,1)(12,7)
\psline{-}(13,1)(13,7) \psline{-}(14,1)(14,7) \psline{-}(15,1)(15,7)
\psline{-}(16,1)(16,7) \psline{-}(17,1)(17,7) \psline{-}(18,1)(18,7)
\psline{-}(19,1)(19,7) \psline{-}(20,1)(20,7) \psline{-}(21,1)(21,7)
\psline{-}(22,1)(22,7) \psline{-}(23,1)(23,7) \psline{-}(24,1)(24,7)
\psline{-}(25,1)(25,7) \psline{-}(26,1)(26,7) \psline{-}(27,1)(27,7)
\psline{-}(28,1)(28,7) \psline{-}(29,1)(29,7) \psline{-}(30,1)(30,7)
\psline{-}(31,1)(31,7) \psline{-}(32,1)(32,7) \psline{-}(33,1)(33,7)
\psline{-}(34,1)(34,7) \psline{-}(35,1)(35,7) \psline{-}(36,1)(36,7)
\psline{-}(37,1)(37,7) \psline{-}(38,1)(38,7) \psline{-}(39,1)(39,7)
\psline{-}(40,1)(40,7) \psline{-}(41,1)(41,7) \psline{-}(42,1)(42,7)
\psline{-}(43,1)(43,7) \psline{-}(44,1)(44,7) \psline{-}(45,1)(45,7)
\psline{-}(46,1)(46,7)
%axes
\psset{linewidth=0.25pt,fillstyle=none,linestyle=solid,linecolor=black}
\psline{->}(0,1)(46.5,1)
\psline{-}(0,7)(46,7)
\psline{-}(0,1)(0,7)
%numeros
\rput(-0.5,1){\scriptsize $1$}\rput(-0.5,3){\scriptsize $2$}
\rput(-0.5,5){\scriptsize $3$}\rput(-0.5,7){\scriptsize $4$}
\rput(0,0.5){\scriptsize $0$} \rput(2,0.5){\scriptsize $1$}
\rput(4,0.5){\scriptsize $2$} \rput(6,0.5){\scriptsize $3$}
\rput(8,0.5){\scriptsize $4$} \rput(10,0.5){\scriptsize $5$}
\rput(12,0.5){\scriptsize $6$} \rput(14,0.5){\scriptsize $7$}
\rput(16,0.5){\scriptsize $8$} \rput(18,0.5){\scriptsize $9$}
\rput(20,0.5){\scriptsize $10$} \rput(22,0.5){\scriptsize $11$}
\rput(24,0.5){\scriptsize $12$} \rput(26,0.5){\scriptsize $13$}
\rput(28,0.5){\scriptsize $14$} \rput(30,0.5){\scriptsize $15$}
\rput(32,0.5){\scriptsize $16$} \rput(34,0.5){\scriptsize $17$}
\rput(36,0.5){\scriptsize $18$} \rput(38,0.5){\scriptsize $19$}
\rput(40,0.5){\scriptsize $20$} \rput(42,0.5){\scriptsize $21$}
\rput(44,0.5){\scriptsize $22$} \rput(46,0.5){\scriptsize $23$}
%path
\psset{linewidth=0.7pt,fillstyle=none,linestyle=solid,linecolor=black}
\psline(0,7)(1,6)(2,5)(3,4)(4,3)(5,2)(6,1)(7,2)(8,1)(9,2)(10,3)(11,4)
 (12,3)(13,4)(14,5)(15,6)(16,5)(17,6)(18,5)(19,6)(20,5)(21,6)(22,5)
 (23,6)(24,5)(25,6)(26,5)(27,6)(28,5)(29,6)(30,5)(31,6)(32,5)(33,6)
 (34,5)(35,6)(36,5)(37,6)(38,5)(39,6)(40,5)(41,6)(42,5)(43,6)(44,5)
 (45,6)(46,5)
\end{pspicture}
\end{center}
\end{figure}
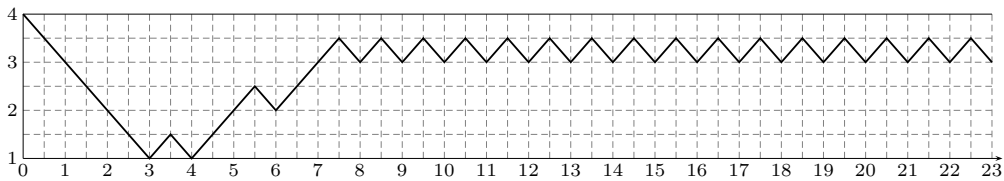

Now obtain the path $\hlac\in\HH^p_{a/2,b/2}$
by shrinking the path $\hcut$ by a factor of two,
and discarding the lowermost band, from which the
path is necessarily absent.
From $\hcut$, we thus obtain the path $\hlac$ given in
Fig.~\ref{TypicalScoringHat}.

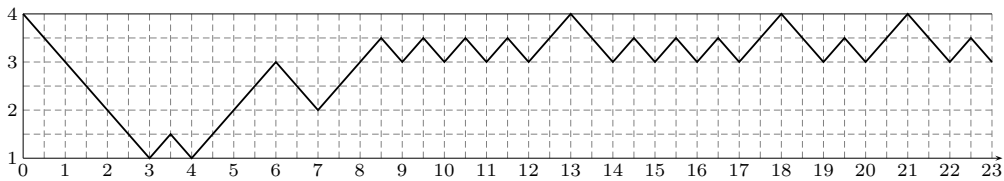
\begin{figure}[ht]
\caption{Half-lattice path $\hh\in\mathcal{H}_{4,3}^{4}$.}
\vskip-1mm
\label{TypicalHalfHat}
\begin{center}
\psset{yunit=0.32cm,xunit=0.28cm}
\begin{pspicture}(0,-0.5)(46,8)
%dashed grid
\psset{linewidth=0.25pt,linestyle=dashed, dash=2.5pt 1.5pt,linecolor=gray}
\psline{-}(0,2)(46,2) \psline{-}(0,3)(46,3)
\psline{-}(0,4)(46,4) \psline{-}(0,5)(46,5) \psline{-}(0,6)(46,6)
\psline{-}(1,1)(1,7) \psline{-}(2,1)(2,7) \psline{-}(3,1)(3,7)
\psline{-}(4,1)(4,7) \psline{-}(5,1)(5,7) \psline{-}(6,1)(6,7)
\psline{-}(7,1)(7,7) \psline{-}(8,1)(8,7) \psline{-}(9,1)(9,7)
\psline{-}(10,1)(10,7) \psline{-}(11,1)(11,7) \psline{-}(12,1)(12,7)
\psline{-}(13,1)(13,7) \psline{-}(14,1)(14,7) \psline{-}(15,1)(15,7)
\psline{-}(16,1)(16,7) \psline{-}(17,1)(17,7) \psline{-}(18,1)(18,7)
\psline{-}(19,1)(19,7) \psline{-}(20,1)(20,7) \psline{-}(21,1)(21,7)
\psline{-}(22,1)(22,7) \psline{-}(23,1)(23,7) \psline{-}(24,1)(24,7)
\psline{-}(25,1)(25,7) \psline{-}(26,1)(26,7) \psline{-}(27,1)(27,7)
\psline{-}(28,1)(28,7) \psline{-}(29,1)(29,7) \psline{-}(30,1)(30,7)
\psline{-}(31,1)(31,7) \psline{-}(32,1)(32,7) \psline{-}(33,1)(33,7)
\psline{-}(34,1)(34,7) \psline{-}(35,1)(35,7) \psline{-}(36,1)(36,7)
\psline{-}(37,1)(37,7) \psline{-}(38,1)(38,7) \psline{-}(39,1)(39,7)
\psline{-}(40,1)(40,7) \psline{-}(41,1)(41,7) \psline{-}(42,1)(42,7)
\psline{-}(43,1)(43,7) \psline{-}(44,1)(44,7) \psline{-}(45,1)(45,7)
\psline{-}(46,1)(46,7)
%axes
\psset{linewidth=0.25pt,fillstyle=none,linestyle=solid,linecolor=black}
\psline{->}(0,1)(46.5,1)
\psline{-}(0,7)(46,7)
\psline{-}(0,1)(0,7)
%numeros
\rput(-0.5,1){\scriptsize $1$}\rput(-0.5,3){\scriptsize $2$}
\rput(-0.5,5){\scriptsize $3$}\rput(-0.5,7){\scriptsize $4$}
\rput(0,0.5){\scriptsize $0$} \rput(2,0.5){\scriptsize $1$}
\rput(4,0.5){\scriptsize $2$} \rput(6,0.5){\scriptsize $3$}
\rput(8,0.5){\scriptsize $4$} \rput(10,0.5){\scriptsize $5$}
\rput(12,0.5){\scriptsize $6$} \rput(14,0.5){\scriptsize $7$}
\rput(16,0.5){\scriptsize $8$} \rput(18,0.5){\scriptsize $9$}
\rput(20,0.5){\scriptsize $10$} \rput(22,0.5){\scriptsize $11$}
\rput(24,0.5){\scriptsize $12$} \rput(26,0.5){\scriptsize $13$}
\rput(28,0.5){\scriptsize $14$} \rput(30,0.5){\scriptsize $15$}
\rput(32,0.5){\scriptsize $16$} \rput(34,0.5){\scriptsize $17$}
\rput(36,0.5){\scriptsize $18$} \rput(38,0.5){\scriptsize $19$}
\rput(40,0.5){\scriptsize $20$} \rput(42,0.5){\scriptsize $21$}
\rput(44,0.5){\scriptsize $22$} \rput(46,0.5){\scriptsize $23$}
%path
\psset{linewidth=0.7pt,fillstyle=none,linestyle=solid,linecolor=black}
\psline(0,7)(1,6)(2,5)(3,4)(4,3)(5,2)(6,1)(7,2)(8,1)(9,2)(10,3)(11,4)
 (12,5)(13,4)(14,3)(15,4)(16,5)(17,6)(18,5)(19,6)(20,5)(21,6)(22,5)
 (23,6)(24,5)(25,6)(26,7)(27,6)(28,5)(29,6)(30,5)(31,6)(32,5)(33,6)
 (34,5)(35,6)(36,7)(37,6)(38,5)(39,6)(40,5)(41,6)(42,7)(43,6)(44,5)
 (45,6)(46,5)
\end{pspicture}
\end{center}
\end{figure}

Finally, we obtain the bijective image $\hh$ of the original $h$ upon
heightening some of the non-integer height peaks of $\hlac$,
by in each case inserting a NE-SE pair of edges.
After setting
\begin{equation}
\label{Eq:StaggerParts}
\mu_i=\lambda_i+n+1-i\qquad\text{for}\qquad 1\le i\le n,
\end{equation}
the peaks of $\hlac$ to be heightened are those
numbered $\mu_1,\mu_2,\ldots,\mu_n$ from the left.
In the ongoing example, $\mu_1=13$, $\mu_2=11$, $\mu_3=7$ and $\mu_4=2$,
thereby leading to the half-lattice path $\hh$
of Fig.~\ref{TypicalHalfHat}.
It may be checked that, in this case, $\unwt(\hh)=297/4$ and
$\unwt(\hhgs)=1/4$ so that $\wh(\hh)=74$, which equals 
the weight $\wt(h)$ of Fig.~\ref{TypicalScoringFig}, as required.

This completes the description of the
bijection~\eqref{Eq:Bijection1}.

\subsection{Proof of $\M(p,2p+1)$ bijection}

In what follows, we show that the combined map
\begin{equation}
\label{Eq:CombBij}
h\to(\hcut,n,\lambda)\to(\hlac,n,\mu)\to\hat h
\end{equation}
is a weight-preserving bijection between
$\PP^{p,2p+1}_{a,b}$ and $\HH^p_{\ha,\hb}$
where we set $\ha=a/2$, $\hb=b/2$,
$\lambda=(\lambda_1,\lambda_2,\ldots,\lambda_n)$ and
$\mu=(\mu_1,\mu_2,\ldots,\mu_n)$.
Note that $\lambda$ is a partition with at most $n$ parts,
and $\mu$, as defined by \eqref{Eq:StaggerParts}, is a distinct partition
with exactly $n$ parts.

That it is a bijection follows because the inverse map from
$\HH^p_{\ha,\hb}$ to $\PP^{p,2p+1}_{a,b}$,
which is easily described, is well-defined.
This relies on the fact that each $\hh\in\HH^p_{\ha,\hb}$
arises from a unique half-lattice path, $\hlac$, whose characteristic
property, we recall, is that its peaks are all at
non-integer heights;
$\hlac$ is thus recovered from $\hat h$ by shrinking each integer peak.
With $n$ the number of such integer peaks,
the partition $\mu=(\mu_1,\ldots,\mu_n)$ is determined by
setting its parts to be the numberings of the integer peaks
amongst all peaks, counted from the left.
The parts of $\mu$ are necessarily distinct, and thus
a genuine partition $\lambda$ is recovered via
\eqref{Eq:StaggerParts}.
%That the map $h\to\hat h$ is weight-preserving is demonstrated below.

\subsection{Proof of $\M(p,2p+1)$ weight preservation (i)}
\label{Sec:Proof1_1}

The RSOS path $\hcut$ is obtained from $h$ by removing the
particles numbered $\lambda_1,\lambda_2,\ldots,\lambda_n$.
Consider the removal of the $i$th such particle from $h$.
The number of non-scoring vertices to its left is $\lambda_i$.
Let $k$ be the total number of scoring vertices in $h$.
The band structure for the $p'=2p+1$ cases
ensures that the first of the two scoring vertices that
comprise the particle is necessarily a peak or valley.
Let it be at position $(x,h_x)$.
If it is a peak, then the following vertex is at $(x+1,h_x-1)$,
and together they contribute
\begin{equation}
u_x+v_{x+1}=\frac12(x-h_x+a+x+1+h_x-1-a)=x
\end{equation}
to $\wt(h)$.
If it is a valley, then the following vertex is at $(x+1,h_x+1)$,
and together they contribute
\begin{equation}
v_x+u_{x+1}=\frac12(x+h_x-a+x+1-h_x-1+a)=x
\end{equation}
to $\wt(h)$.
There are $x-1-\lambda_i$ scoring vertices to the left of
the particle, and thus $\lambda_i+k-x-1$ to its right.
On removing the particle, the contribution of each of the latter
to the weight decreases by one.
Thus the total weight reduction on removing the particle is
$\lambda_i+k-1$.

%Then, on removing all $n$ particles,
Then, on removing all $n$ of the particles from $h$,
noting that $k$ decreases by two at each step, we obtain
\begin{equation}\label{Eq:RemFB}
\begin{split}
\wt(h^{\mathrm{cut}})
&=\wt(h) - \sum_{i=1}^n \lambda_i - (k-1)-(k-3)-(k-5)-\cdots-(k-2n-1)\\
&=\wt(h) - \sum_{i=1}^n \lambda_i - n(k-n).
\end{split}
\end{equation}

\subsection{Proof of $\M(p,2p+1)$ weight preservation (ii)}
\label{Sec:Proof1_2}

The path $\hcut\in\PP^{p,2p+1}_{a,b}$ has no neighbouring pair of
scoring vertices,
which, as is easily seen, implies it has
no valleys at odd height, and no peaks at even height.
The half-lattice path $\hlac\in\HH^p_{\ha,\hb}$, obtained by
shrinking $\hcut$ by a factor of 2, then has no valleys at
non-integer height, and no peaks at integer height.
In this section, we show that $\wh(\hlac)=\wt(\hcut)$.

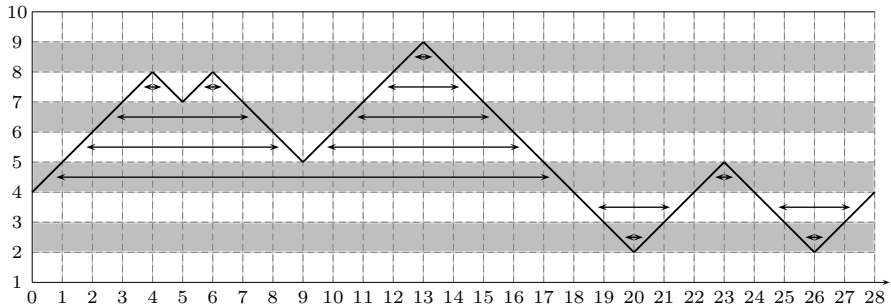
\begin{figure}[ht]
\caption{Example of segment matching for a path $h$.}
%\vskip-1.2cm
\label{TypicalMatching}
\vskip-2.2cm
\begin{center}
\psset{unit=0.4cm}
%\begin{pspicture}(1,0)(11,6.5)
\begin{pspicture}(0,-0.5)(28,11)
%dark bands
\psset{linestyle=none,fillstyle=solid,fillcolor=lightgray}
\psframe(0,2)(28,3)
\psframe(0,4)(28,5)
\psframe(0,6)(28,7)
\psframe(0,8)(28,9)
%dashed grid
\psset{linewidth=0.25pt,linestyle=dashed, dash=2.5pt 1.5pt,linecolor=gray}
\psline{-}(0,2)(28,2) \psline{-}(0,3)(28,3) \psline{-}(0,4)(28,4)
\psline{-}(0,5)(28,5) \psline{-}(0,6)(28,6) \psline{-}(0,7)(28,7)
\psline{-}(0,8)(28,8) \psline{-}(0,9)(28,9)
\psline{-}(1,1)(1,10) \psline{-}(2,1)(2,10) \psline{-}(3,1)(3,10)
\psline{-}(4,1)(4,10) \psline{-}(5,1)(5,10) \psline{-}(6,1)(6,10)
\psline{-}(7,1)(7,10) \psline{-}(8,1)(8,10) \psline{-}(9,1)(9,10)
\psline{-}(10,1)(10,10) \psline{-}(11,1)(11,10) \psline{-}(12,1)(12,10)
\psline{-}(13,1)(13,10) \psline{-}(14,1)(14,10) \psline{-}(15,1)(15,10)
\psline{-}(16,1)(16,10) \psline{-}(17,1)(17,10) \psline{-}(18,1)(18,10)
\psline{-}(19,1)(19,10) \psline{-}(20,1)(20,10) \psline{-}(21,1)(21,10)
\psline{-}(22,1)(22,10) \psline{-}(23,1)(23,10) \psline{-}(24,1)(24,10)
\psline{-}(25,1)(25,10) \psline{-}(26,1)(26,10) \psline{-}(27,1)(27,10)
\psline{-}(28,1)(28,10)
%axes
\psset{linewidth=0.25pt,fillstyle=none,linestyle=solid,linecolor=black}
\psline{->}(0,1)(28.5,1)
\psline{-}(0,10)(28,10)
\psline{-}(0,1)(0,10)
%numeros
\rput(-0.5,1){\scriptsize $1$}\rput(-0.5,2){\scriptsize $2$}
\rput(-0.5,3){\scriptsize $3$}\rput(-0.5,4){\scriptsize $4$}
\rput(-0.5,5){\scriptsize $5$}\rput(-0.5,6){\scriptsize $6$}
\rput(-0.5,7){\scriptsize $7$}\rput(-0.5,8){\scriptsize $8$}
\rput(-0.5,9){\scriptsize $9$}\rput(-0.5,10){\scriptsize $10$}
\rput(0,0.5){\scriptsize $0$} \rput(1,0.5){\scriptsize $1$}
\rput(2,0.5){\scriptsize $2$} \rput(3,0.5){\scriptsize $3$}
\rput(4,0.5){\scriptsize $4$} \rput(5,0.5){\scriptsize $5$}
\rput(6,0.5){\scriptsize $6$} \rput(7,0.5){\scriptsize $7$}
\rput(8,0.5){\scriptsize $8$} \rput(9,0.5){\scriptsize $9$}
\rput(10,0.5){\scriptsize $10$} \rput(11,0.5){\scriptsize $11$}
\rput(12,0.5){\scriptsize $12$} \rput(13,0.5){\scriptsize $13$}
\rput(14,0.5){\scriptsize $14$} \rput(15,0.5){\scriptsize $15$}
\rput(16,0.5){\scriptsize $16$} \rput(17,0.5){\scriptsize $17$}
\rput(18,0.5){\scriptsize $18$} \rput(19,0.5){\scriptsize $19$}
\rput(20,0.5){\scriptsize $20$} \rput(21,0.5){\scriptsize $21$}
\rput(22,0.5){\scriptsize $22$} \rput(23,0.5){\scriptsize $23$}
\rput(24,0.5){\scriptsize $24$} \rput(25,0.5){\scriptsize $25$}
\rput(26,0.5){\scriptsize $26$} \rput(27,0.5){\scriptsize $27$}
\rput(28,0.5){\scriptsize $28$}
%path
\psset{linewidth=0.7pt,fillstyle=none,linestyle=solid,linecolor=black}
\psline{-}(0,4)(4,8) 
\psline{-}(4,8)(5,7) \psline{-}(5,7)(6,8)
\psline{-}(6,8)(9,5) \psline{-}(9,5)(13,9)
\psline{-}(13,9)(20,2) \psline{-}(20,2)(23,5)
\psline{-}(23,5)(26,2) \psline{-}(26,2)(28,4)
%arrows
\psset{linewidth=0.5pt,linestyle=solid,linecolor=black}
\psline{<->}(0.8,4.5)(17.2,4.5)
\psline{<->}(1.8,5.5)(8.2,5.5)
\psline{<->}(9.8,5.5)(16.2,5.5)
\psline{<->}(2.8,6.5)(7.2,6.5)
\psline{<->}(10.8,6.5)(15.2,6.5)
\psline{<->}(3.7,7.5)(4.3,7.5)
\psline{<->}(5.7,7.5)(6.3,7.5)
\psline{<->}(11.8,7.5)(14.2,7.5)
\psline{<->}(12.7,8.5)(13.3,8.5)
\psline{<->}(22.7,4.5)(23.3,4.5)
\psline{<->}(18.8,3.5)(21.2,3.5)
\psline{<->}(19.7,2.5)(20.3,2.5)
\psline{<->}(24.8,3.5)(27.2,3.5)
\psline{<->}(25.7,2.5)(26.3,2.5)
%\psline{<->}(22.8,4.5)(25.2,4.5)
%\psline{<->}(23.7,3.5)(24.3,3.5)
%
\end{pspicture}
\end{center}
\end{figure}

First consider the case where $a=b$.
On truncating the path $\hcut$ after $L(\hcut)$ steps,
the path then starts and finishes at the same height.
We may then match (pair) each path segment with another
at the same height, one NE and one SE (the order is immaterial).
This matching process is illustrated in Fig.~\ref{TypicalMatching}.

The above restrictions on the peaks and valleys of $\hcut$
imply that the scoring vertices of $\hcut$ occur
at the right ends of all the segments in the light bands.
Consider a specific matched pair of segments
in a light band,
with the left ends of the
NE and SE segments at positions $(x,y)$ and $(x',y+1)$ respectively.
The two scoring vertices at $(x+1,y+1)$ and $(x'+1,y)$ together contribute
\begin{equation}
\label{Eq:RSOS2Cont}
u_{x+1}+v_{x'+1}=\frac12(x+1-(y+1)+a+x'+1+y-a)=\frac12(x+x'+1)
\end{equation}
to $\wt(\hcut)$.

For the corresponding half-lattice path $\hlac$,
consider the corresponding matched pair of segments.
The four straight vertices at the ends of each of these segments
contribute
\begin{equation}
\frac12\left[\frac12x+\frac12(x+1)+\frac12x'+\frac12(x'+1)\right]
=\frac12(x+x'+1)
\end{equation}
to $\unwt(\hlac)$
(the vertex at $(0,\ha)$ will be required here if it is straight).
Since this agrees with the contribution \eqref{Eq:RSOS2Cont} of the
corresponding two scoring vertices of $\hcut$ to
$\wt(\hcut)$, we conclude that $\unwt(\hlac)=\wt(\hcut)$.
But, for $a=b$, $\wh(\hlac)=\unwt(\hlac)$,
thereby proving that $\wh(\hlac)=\wt(\hcut)$ in this $a=b$ case.

In the case that $a\ne b$, we make use of the trick described in
Section \ref{Sec:WtTrick} to obtain $\wh(\hlac)$ by extending the
path $\hlac$ to the left.
On the other hand,
extending the path $\hcut$ to the left by $2e=\vert a-b\vert$ steps
with $\hcut_{-2e}=b$, creates an RSOS path of unchanged weight $\wt(\hcut)$
because, via \eqref{Eq:ScoreWts}, the additional scoring vertices
each contribute 0 to the weight.
Then, upon applying the argument used above in the $a=b$ case to
these extended paths, we obtain $\wh(\hlac)=\wt(\hcut)$ for
$a\ne b$ also.

\subsection{Proof of $\M(p,2p+1)$ weight preservation (iii)}
\label{Sec:Proof1_3}

Consider a half-lattice path
$\hh^{(0)}\in\HH^p_{\ha,\hb}$,
having $\numstr$ straight vertices. % not segments.
This count includes consideration of the vertex at $(0,\ha)$,
which, through the convention stated in Section \ref{Sec:Spec},
is deemed straight if and only if the first segment of $\hh^{(0)}$
is in the SE direction.
The vertices of $\hh^{(0)}$ that do not contribute to
its weight are the peaks and valleys.
We now determine the change in weight on heightening one of the peaks.
Let the peak being heightened be the $j$th, counting from the left,
and let it be situated at $(x,\hh^{(0)}_x)$.
There are necessarily $j$ valleys to the left of this peak
(perhaps including one at $(0,\ha)$), and therefore
$2x+1-2j$ straight vertices.
After this position, there are thus $\numstr-2x+2j-1$ straight vertices.
The heightening moves each of these to the right by two
(half-integer) positions.
It also introduces two straight vertices, at positions
$(x,\hh^{(0)}_x)$ and $(x+1,\hh^{(0)}_x)$.
Thus, if this resulting path is denoted $\hh^{(1)}$,
\begin{equation}
\begin{split}
\wh(\hh^{(1)})
&=\wh(\hh^{(0)})
+\frac{1}{2} (x+(x+1))+\frac12(\numstr-2x+2j-1))\\
&=\wh(\hh^{(0)}) +\frac{1}{2}\numstr+j.
\end{split}
\end{equation}

Note that the heightening increases the value of $\numstr$ by 2.
So if we obtain the path $\hh$ by performing a succession of
heightenings to a path $\hlac$ at peaks numbered
$\mu_1$, $\mu_2$, $\ldots,\mu_n$,
we have
\begin{equation}\label{Eq:DeeperVal}
\begin{split}
\wh(\hh)
&=\wh(\hlac)
+\frac{1}{2} (\numstr+(\numstr+2)+\cdots+(\numstr+2n-2))
+\sum_{i=1}^n \mu_i\\
&=\wh(\hlac)
+\frac{n}{2} (\numstr+n-1) + \sum_{i=1}^n \mu_i.
\end{split}
\end{equation}

\subsection{Proof of $\M(p,2p+1)$ weight preservation (iv)}
\label{Sec:Proof1_4}

Consider the combined map \eqref{Eq:CombBij}.
Let $k$ and $k'$ be the number of
scoring vertices in $h$ and $\hcut$ respectively,
and let $\numstr$ be the number of straight vertices.
%(including consideration of the vertex at $(0,\hat a)$) in $\hlac$.
The pair removal process in Section \ref{Sec:Proof1_1} shows
that $k'=k-2n$.
The matching of edges described in Section \ref{Sec:Proof1_3}
shows that $\numstr=2k'$ and thus $\numstr=2k-4n$.
Then, using \eqref{Eq:RemFB} and \eqref{Eq:DeeperVal}, and
the fact that $\wh(\hlac)=\wt(\hcut)$, we obtain
\begin{equation}\label{Eq:WtDiff}
\begin{split}
\wh(\hh)-\wt(h)
&=\sum_{i=1}^n \mu_i -\sum_{i=1}^n \lambda_i
+\frac n2(\numstr+n-1)-n(k-n)\\
&=\sum_{i=1}^n \mu_i -\sum_{i=1}^n \lambda_i
-\frac n2(n+1)\\
&=0,
\end{split}
\end{equation}
where the final equality follows from \eqref{Eq:StaggerParts} because
$\sum_{i=1}^n\mu_i-\sum_{i=1}^n\lambda_i
=\sum_{i=1}^n i=\frac 12n(n+1)$.
Thus the weight-preserving nature of the bijection
\eqref{Eq:CombBij}
has been proved.

\section{The $\M(p,2p-1)$ bijection}
\label{Sec:Bijection2}

\subsection{Description}

Here, we describe a weight-preserving bijection
\begin{equation}\label{Eq:Bijection2}
\PP^{p,2p-1}_{a,b-1}\leftrightarrow \HH^{p-1/2}_{b/2,a/2} ,
\end{equation}
for $p\in\ZZ_{\ge3}$, and $a$ and $b$ even integers with
$1<a,b<2p-1$.
From this, via \eqref{Eq:XGFdef}, \eqref{Eq:XRocha} and \eqref{Eq:YGFdef},
the identity \eqref{Eq:HalfChar2} is immediately established.

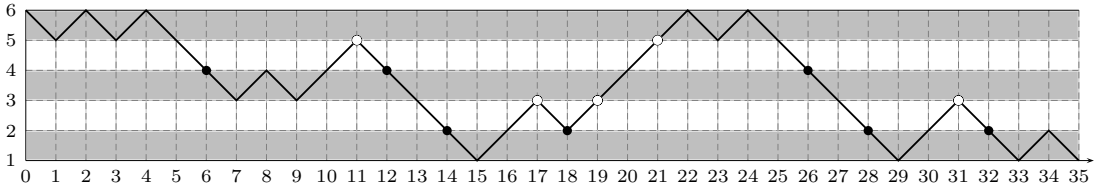
\begin{figure}[ht]
\caption{An RSOS path $h\in \PP^{4,7}_{6,1}$.}
\vskip-0.2cm
\label{TypicalDualFig}
\begin{center}
\psset{unit=0.4cm}
\begin{pspicture}(0,-0.5)(35,7)
%dark bands
\psset{linestyle=none,fillstyle=solid,fillcolor=lightgray}
\psframe(0,1)(35,2)
\psframe(0,3)(35,4)
\psframe(0,5)(35,6)
%dashed grid
\psset{linewidth=0.25pt,linestyle=dashed, dash=2.5pt 1.5pt,linecolor=gray}
\psline{-}(0,2)(35,2) \psline{-}(0,3)(35,3) \psline{-}(0,4)(35,4)
\psline{-}(0,5)(35,5)
\psline{-}(1,1)(1,6) \psline{-}(2,1)(2,6) \psline{-}(3,1)(3,6)
\psline{-}(4,1)(4,6) \psline{-}(5,1)(5,6) \psline{-}(6,1)(6,6)
\psline{-}(7,1)(7,6) \psline{-}(8,1)(8,6) \psline{-}(9,1)(9,6)
\psline{-}(10,1)(10,6) \psline{-}(11,1)(11,6) \psline{-}(12,1)(12,6)
\psline{-}(13,1)(13,6) \psline{-}(14,1)(14,6) \psline{-}(15,1)(15,6)
\psline{-}(16,1)(16,6) \psline{-}(17,1)(17,6) \psline{-}(18,1)(18,6)
\psline{-}(19,1)(19,6) \psline{-}(20,1)(20,6) \psline{-}(21,1)(21,6)
\psline{-}(22,1)(22,6) \psline{-}(23,1)(23,6) \psline{-}(24,1)(24,6)
\psline{-}(25,1)(25,6) \psline{-}(26,1)(26,6) \psline{-}(27,1)(27,6)
\psline{-}(28,1)(28,6) \psline{-}(29,1)(29,6) \psline{-}(30,1)(30,6)
\psline{-}(31,1)(31,6) \psline{-}(32,1)(32,6) \psline{-}(33,1)(33,6)
\psline{-}(34,1)(34,6) \psline{-}(35,1)(35,6)
%axes
\psset{linewidth=0.25pt,fillstyle=none,linestyle=solid,linecolor=black}
\psline{->}(0,1)(35.5,1)
\psline{-}(0,6)(35,6)
\psline{-}(0,1)(0,6)
%numeros
\rput(-0.5,1){\scriptsize $1$}\rput(-0.5,2){\scriptsize $2$}
\rput(-0.5,3){\scriptsize $3$}\rput(-0.5,4){\scriptsize $4$}
\rput(-0.5,5){\scriptsize $5$}\rput(-0.5,6){\scriptsize $6$}
\rput(0,0.5){\scriptsize $0$} \rput(1,0.5){\scriptsize $1$}
\rput(2,0.5){\scriptsize $2$} \rput(3,0.5){\scriptsize $3$}
\rput(4,0.5){\scriptsize $4$} \rput(5,0.5){\scriptsize $5$}
\rput(6,0.5){\scriptsize $6$} \rput(7,0.5){\scriptsize $7$}
\rput(8,0.5){\scriptsize $8$} \rput(9,0.5){\scriptsize $9$}
\rput(10,0.5){\scriptsize $10$} \rput(11,0.5){\scriptsize $11$}
\rput(12,0.5){\scriptsize $12$} \rput(13,0.5){\scriptsize $13$}
\rput(14,0.5){\scriptsize $14$} \rput(15,0.5){\scriptsize $15$}
\rput(16,0.5){\scriptsize $16$} \rput(17,0.5){\scriptsize $17$}
\rput(18,0.5){\scriptsize $18$} \rput(19,0.5){\scriptsize $19$}
\rput(20,0.5){\scriptsize $20$} \rput(21,0.5){\scriptsize $21$}
\rput(22,0.5){\scriptsize $22$} \rput(23,0.5){\scriptsize $23$}
\rput(24,0.5){\scriptsize $24$} \rput(25,0.5){\scriptsize $25$}
\rput(26,0.5){\scriptsize $26$} \rput(27,0.5){\scriptsize $27$}
\rput(28,0.5){\scriptsize $28$} \rput(29,0.5){\scriptsize $29$}
\rput(30,0.5){\scriptsize $30$} \rput(31,0.5){\scriptsize $31$}
\rput(32,0.5){\scriptsize $32$} \rput(33,0.5){\scriptsize $33$}
\rput(34,0.5){\scriptsize $34$} \rput(35,0.5){\scriptsize $35$}
%path
\psset{linewidth=0.7pt,fillstyle=none,linestyle=solid,linecolor=black}
\psline(0,6)(1,5)(2,6)(3,5)(4,6)(5,5)(6,4)(7,3)(8,4)(9,3)(10,4)(11,5)
 (12,4)(13,3)(14,2)(15,1)(16,2)(17,3)(18,2)(19,3)(20,4)(21,5)(22,6)
 (23,5)(24,6)(25,5)(26,4)(27,3)(28,2)(29,1)(30,2)(31,3)(32,2)(33,1)
 (34,2)(35,1)
%dots
\psset{dotsize=3.5pt,fillcolor=black}
\psdots(6,4)(12,4)(14,2)(18,2)(26,4)(28,2)(32,2)
\psset{dotsize=4pt,dotstyle=o,fillcolor=white}
\psdots(11,5)(17,3)(19,3)(21,5)(31,3)
\end{pspicture}
\end{center}
\end{figure}

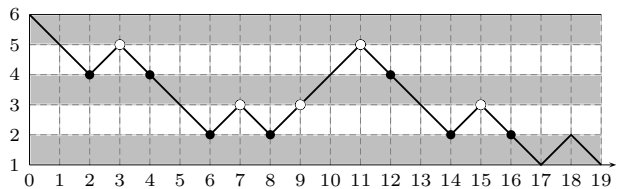
\begin{figure}[ht]
\caption{$\hcut\in \PP^{4,7}_{6,1}$ obtained from
                Fig.~\ref{TypicalDualFig}}
\vskip-0.2cm
\label{TypicalDualCut}
\begin{center}
\psset{unit=0.4cm}
\begin{pspicture}(0,-0.5)(19,7)
%dark bands
\psset{linestyle=none,fillstyle=solid,fillcolor=lightgray}
\psframe(0,1)(19,2)
\psframe(0,3)(19,4)
\psframe(0,5)(19,6)
%dashed grid
\psset{linewidth=0.25pt,linestyle=dashed, dash=2.5pt 1.5pt,linecolor=gray}
\psline{-}(0,2)(19,2) \psline{-}(0,3)(19,3) \psline{-}(0,4)(19,4)
\psline{-}(0,5)(19,5)
\psline{-}(1,1)(1,6) \psline{-}(2,1)(2,6) \psline{-}(3,1)(3,6)
\psline{-}(4,1)(4,6) \psline{-}(5,1)(5,6) \psline{-}(6,1)(6,6)
\psline{-}(7,1)(7,6) \psline{-}(8,1)(8,6) \psline{-}(9,1)(9,6)
\psline{-}(10,1)(10,6) \psline{-}(11,1)(11,6) \psline{-}(12,1)(12,6)
\psline{-}(13,1)(13,6) \psline{-}(14,1)(14,6) \psline{-}(15,1)(15,6)
\psline{-}(16,1)(16,6) \psline{-}(17,1)(17,6) \psline{-}(18,1)(18,6)
\psline{-}(19,1)(19,6)
%axes
\psset{linewidth=0.25pt,fillstyle=none,linestyle=solid,linecolor=black}
\psline{->}(0,1)(19.5,1)
\psline{-}(0,6)(19,6)
\psline{-}(0,1)(0,6)
%numeros
\rput(-0.5,1){\scriptsize $1$}\rput(-0.5,2){\scriptsize $2$}
\rput(-0.5,3){\scriptsize $3$}\rput(-0.5,4){\scriptsize $4$}
\rput(-0.5,5){\scriptsize $5$}\rput(-0.5,6){\scriptsize $6$}
\rput(0,0.5){\scriptsize $0$} \rput(1,0.5){\scriptsize $1$}
\rput(2,0.5){\scriptsize $2$} \rput(3,0.5){\scriptsize $3$}
\rput(4,0.5){\scriptsize $4$} \rput(5,0.5){\scriptsize $5$}
\rput(6,0.5){\scriptsize $6$} \rput(7,0.5){\scriptsize $7$}
\rput(8,0.5){\scriptsize $8$} \rput(9,0.5){\scriptsize $9$}
\rput(10,0.5){\scriptsize $10$} \rput(11,0.5){\scriptsize $11$}
\rput(12,0.5){\scriptsize $12$} \rput(13,0.5){\scriptsize $13$}
\rput(14,0.5){\scriptsize $14$} \rput(15,0.5){\scriptsize $15$}
\rput(16,0.5){\scriptsize $16$} \rput(17,0.5){\scriptsize $17$}
\rput(18,0.5){\scriptsize $18$} \rput(19,0.5){\scriptsize $19$}
%path
\psset{linewidth=0.7pt,fillstyle=none,linestyle=solid,linecolor=black}
\psline(0,6)(1,5)(2,4)(3,5)(4,4)(5,3)(6,2)(7,3)(8,2)(9,3)(10,4)(11,5)
 (12,4)(13,3)(14,2)(15,3)(16,2)(17,1)(18,2)(19,1)
%dots
\psset{dotsize=3.5pt,fillcolor=black}
\psdots(2,4)(4,4)(6,2)(8,2)(12,4)(14,2)(16,2)
\psset{dotsize=4pt,dotstyle=o,fillcolor=white}
\psdots(3,5)(7,3)(9,3)(11,5)(15,3)
\end{pspicture}
\end{center}
\end{figure}

Let $h\in\PP^{p,2p-1}_{a,b-1}$, and note that in this case, the
RSOS band structure has alternating white and dark bands,
with the uppermost and lowermost bands both dark.
In this case, particles are identified with pairs of
adjacent \emph{non-scoring} vertices \cite{FLPW}.
For the $i$th particle of $h$, counting from the left,
let $\lambda_i$ be the number of scoring vertices to its right.
Due to the infinite tail, $h$ has an infinite number of
such particles with $\lambda_i=0$ (they form a ``Dirac sea'').
Let $n$ be the largest value for which $\lambda_n>0$.
Then $\lambda=(\lambda_1,\lambda_2,\ldots,\lambda_n)$ is a partition
with exactly $n$ parts.
Now create the path $\hcut\in\PP^{p,2p-1}_{a,b-1}$ by removing
the first $n$ particles from $h$, in each case joining up the loose ends.
To illustrate this, consider the path $h$ given in
Fig.~\ref{TypicalDualFig}, for which we find $\wt(h)=112$.
Here, $n=8$ and $\lambda=(12,12,11,11,8,4,4,2)$.
The removal of these first 8 particles yields
the path $\hcut$ given in Fig.~\ref{TypicalDualCut}.

Let $k$ be the total number of scoring vertices in $\hcut$ (or $h$),
with $m$ the number of those that are peaks, and define $c$
to be the largest value such that $\lambda_c-c\ge k-m$.
Then define the partition $\mu=(\mu_1,\mu_2,\ldots,\mu_c)$
by setting
\begin{equation}
\label{Eq:StaggerParts2}
\mu_i=\lambda_i-i-k+m+1\qquad\text{for}\qquad 1\le i\le c,
\end{equation}
and the partition $\nu=(\nu_1,\nu_2,\ldots,\nu_d)$
for $d=n-c$ by setting
\begin{equation}
\label{Eq:StaggerShift}
\nu_i=\lambda_{i+c}\qquad\text{for}\qquad 1\le i\le d.
\end{equation}
In the case of the path in Fig.~\ref{TypicalDualFig}, we obtain
$k=12$, $m=4$, $c=3$, $\mu=(4,3,1)$, $d=5$ and $\nu=(11,8,4,4,2)$.

We then create a half-lattice path $\hh\in\HH^{p-1/2}_{b/2,a/2}$
from $\hcut$ in three stages.
Since $a,b$ and $L(\hcut)$ are even, it follows that $\hcut_{L(\hcut)}=b$.
Now form a half-lattice path $\hlac\in\HH^{p-1/2}_{b/2,a/2}$
by truncating $\hcut$ after $L(\hcut)$ steps,
flipping it horizontally, shrinking it by a factor of two,
heightening each peak by $1/2$ unit by inserting a NE-SE edge pair,
and finally appending an infinite $(a/2)$-tail.
To obtain the band structure for this $t=p-1/2$ case,
we discard the lowermost band that results from this procedure,
which is between heights $1/2$ and $1$
($\hlac$ doesn't venture there),
and append a band between heights $p-1$ and $p-1/2$ at the top.

In the case of the path $\hcut$ of Fig.~\ref{TypicalDualCut},
note that $L(\hcut)=16$, whereupon this process
yields the path $\hlac$ given in Fig.~\ref{TypicalDualLac}.

\begin{figure}[ht]
\caption{Half-lattice path $\hlac\in\HH^{7/2}_{1,3}$
  obtained from Fig.~\ref{TypicalDualCut}}
\vskip-1mm
\label{TypicalDualLac}
\begin{center}
\psset{yunit=0.32cm,xunit=0.28cm}
\begin{pspicture}(0,-0.5)(30,7)
%dashed grid
\psset{linewidth=0.25pt,linestyle=dashed, dash=2.5pt 1.5pt,linecolor=gray}
\psline{-}(0,2)(30,2) \psline{-}(0,3)(30,3) \psline{-}(0,4)(30,4)
\psline{-}(0,5)(30,5)
\psline{-}(1,1)(1,6) \psline{-}(2,1)(2,6) \psline{-}(3,1)(3,6)
\psline{-}(4,1)(4,6) \psline{-}(5,1)(5,6) \psline{-}(6,1)(6,6)
\psline{-}(7,1)(7,6) \psline{-}(8,1)(8,6) \psline{-}(9,1)(9,6)
\psline{-}(10,1)(10,6) \psline{-}(11,1)(11,6) \psline{-}(12,1)(12,6)
\psline{-}(13,1)(13,6) \psline{-}(14,1)(14,6) \psline{-}(15,1)(15,6)
\psline{-}(16,1)(16,6) \psline{-}(17,1)(17,6) \psline{-}(18,1)(18,6)
\psline{-}(19,1)(19,6) \psline{-}(20,1)(20,6) \psline{-}(21,1)(21,6)
\psline{-}(22,1)(22,6) \psline{-}(23,1)(23,6) \psline{-}(24,1)(24,6)
\psline{-}(25,1)(25,6) \psline{-}(26,1)(26,6) \psline{-}(27,1)(27,6)
\psline{-}(28,1)(28,6) \psline{-}(29,1)(29,6) \psline{-}(30,1)(30,6)
%axes
\psset{linewidth=0.25pt,fillstyle=none,linestyle=solid,linecolor=black}
\psline{->}(0,1)(30.5,1)
\psline{-}(0,6)(30,6)
\psline{-}(0,1)(0,6)
%numeros
\rput(-0.5,1){\scriptsize $1$}\rput(-0.5,3){\scriptsize $2$}
\rput(-0.5,5){\scriptsize $3$}
%\rput(-0.5,6){\scriptsize $9$}
\rput(0,0.3){\scriptsize $0$} \rput(2,0.3){\scriptsize $1$}
\rput(4,0.3){\scriptsize $2$} \rput(6,0.3){\scriptsize $3$}
\rput(8,0.3){\scriptsize $4$} \rput(10,0.3){\scriptsize $5$}
\rput(12,0.3){\scriptsize $6$} \rput(14,0.3){\scriptsize $7$}
\rput(16,0.3){\scriptsize $8$} \rput(18,0.3){\scriptsize $9$}
\rput(20,0.3){\scriptsize $10$} \rput(22,0.3){\scriptsize $11$}
\rput(24,0.3){\scriptsize $12$} \rput(26,0.3){\scriptsize $13$}
\rput(28,0.3){\scriptsize $14$} \rput(30,0.3){\scriptsize $15$}
%path
\psset{linewidth=0.7pt,fillstyle=none,linestyle=solid,linecolor=black}
\psline(0,1)(1,2)(2,3)(3,2)(4,1)(5,2)(6,3)(7,4)(8,5)(9,4)
 (10,3)(11,2)(12,1)(13,2)(14,3)(15,2)(16,1)(17,2)(18,3)
 (19,4)(20,5)(21,4)(22,3)(23,4)(24,5)(25,6)(26,5)(27,6)
 (28,5)(29,6)(30,5)
%\psset{dotsize=4.5pt}\psset{dotstyle=o}
%\psdots(0,1)(4,3)(6,1)(8,3)(12,3)(14,1)(18,3)(20,1)(22,3)(26,5)(28,3)
%\psset{dotstyle=+}
%\psdots(0,1)(4,3)(6,1)(8,3)(12,3)(14,1)(18,3)(20,1)(22,3)(26,5)(28,3)
\end{pspicture}
\end{center}
\end{figure}

The half-lattice path $\hint\in\HH^{p-1/2}_{b/2,a/2}$
is obtained by further heightening some of the peaks
in $\hlac$ by $1/2$ unit by inserting NE-SE edge pairs.
Those that are heightened are the peaks numbered
$\mu_1,\mu_2,\ldots,\mu_c$, counting from the left.

In the case of the path $\hlac$ of Fig.~\ref{TypicalDualLac},
with $\mu=(4,3,1)$,
this yields the path $\hint$ given in Fig.~\ref{TypicalDualInt}.
\begin{figure}[ht]
\caption{Half-lattice path $\hint\in\HH^{7/2}_{1,3}$
  obtained from Fig.~\ref{TypicalDualLac}}
\vskip-1mm
\label{TypicalDualInt}
\begin{center}
\psset{yunit=0.32cm,xunit=0.28cm}
\begin{pspicture}(0,-0.5)(36,7)
%dashed grid
\psset{linewidth=0.25pt,linestyle=dashed, dash=2.5pt 1.5pt,linecolor=gray}
\psline{-}(0,2)(36,2) \psline{-}(0,3)(36,3) \psline{-}(0,4)(36,4)
\psline{-}(0,5)(36,5)
\psline{-}(1,1)(1,6) \psline{-}(2,1)(2,6) \psline{-}(3,1)(3,6)
\psline{-}(4,1)(4,6) \psline{-}(5,1)(5,6) \psline{-}(6,1)(6,6)
\psline{-}(7,1)(7,6) \psline{-}(8,1)(8,6) \psline{-}(9,1)(9,6)
\psline{-}(10,1)(10,6) \psline{-}(11,1)(11,6) \psline{-}(12,1)(12,6)
\psline{-}(13,1)(13,6) \psline{-}(14,1)(14,6) \psline{-}(15,1)(15,6)
\psline{-}(16,1)(16,6) \psline{-}(17,1)(17,6) \psline{-}(18,1)(18,6)
\psline{-}(19,1)(19,6) \psline{-}(20,1)(20,6) \psline{-}(21,1)(21,6)
\psline{-}(22,1)(22,6) \psline{-}(23,1)(23,6) \psline{-}(24,1)(24,6)
\psline{-}(25,1)(25,6) \psline{-}(26,1)(26,6) \psline{-}(27,1)(27,6)
\psline{-}(28,1)(28,6) \psline{-}(29,1)(29,6) \psline{-}(30,1)(30,6)
\psline{-}(31,1)(31,6) \psline{-}(32,1)(32,6) \psline{-}(33,1)(33,6)
\psline{-}(34,1)(34,6) \psline{-}(35,1)(35,6) \psline{-}(36,1)(36,6)
%axes
\psset{linewidth=0.25pt,fillstyle=none,linestyle=solid,linecolor=black}
\psline{->}(0,1)(36.5,1)
\psline{-}(0,6)(36,6)
\psline{-}(0,1)(0,6)
%numeros
\rput(-0.5,1){\scriptsize $1$}\rput(-0.5,3){\scriptsize $2$}
\rput(-0.5,5){\scriptsize $3$}
\rput(0,0.3){\scriptsize $0$} \rput(2,0.3){\scriptsize $1$}
\rput(4,0.3){\scriptsize $2$} \rput(6,0.3){\scriptsize $3$}
\rput(8,0.3){\scriptsize $4$} \rput(10,0.3){\scriptsize $5$}
\rput(12,0.3){\scriptsize $6$} \rput(14,0.3){\scriptsize $7$}
\rput(16,0.3){\scriptsize $8$} \rput(18,0.3){\scriptsize $9$}
\rput(20,0.3){\scriptsize $10$} \rput(22,0.3){\scriptsize $11$}
\rput(24,0.3){\scriptsize $12$} \rput(26,0.3){\scriptsize $13$}
\rput(28,0.3){\scriptsize $14$} \rput(30,0.3){\scriptsize $15$}
\rput(32,0.3){\scriptsize $16$} \rput(34,0.3){\scriptsize $17$}
\rput(36,0.3){\scriptsize $18$}
%path
\psset{linewidth=0.7pt,fillstyle=none,linestyle=solid,linecolor=black}
\psline(0,1)(1,2)(2,3)(3,4)(4,3)(5,2)(6,1)(7,2)(8,3)(9,4)(10,5)(11,4)
 (12,3)(13,2)(14,1)(15,2)(16,3)(17,4)(18,3)(19,2)(20,1)(21,2)(22,3)
 (23,4)(24,5)(25,6)(26,5)(27,4)(28,3)(29,4)(30,5)(31,6)(32,5)(33,6)
 (34,5)(35,6)(36,5)
\psset{dotsize=4.5pt}\psset{dotstyle=o}
\psdots(0,1)(4,3)(6,1)(8,3)(12,3)(14,1)(18,3)(20,1)(22,3)(26,5)(28,3)
\psset{dotstyle=+}
\psdots(0,1)(4,3)(6,1)(8,3)(12,3)(14,1)(18,3)(20,1)(22,3)(26,5)(28,3)
\end{pspicture}
\end{center}
\end{figure}

For $x\in\ZZ_{\ge0}$, the vertex $(x,\hint_x)$ of $\hint$ is designated
an \emph{accretion vertex} if neither $(x,\hint_x)$ nor
$(x+1/2,\hint_{x+1/2})$ is a peak.
We number these from the right.
The path $\hh$ is now obtained from $\hint$ by inserting a number
of NE-SE edge pairs at each accretion vertex:
if $\nu=(\ldots,3^{f_3},2^{f_2},1^{f_1})$ in frequency notation,
$f_j$ such edge pairs are inserted at the $j$th accretion vertex.

\begin{figure}[ht]
\caption{Half-lattice path $\hh\in\HH^{7/2}_{1,3}$ obtained
  from Fig.~\ref{TypicalDualInt}}
\vskip-1mm
\label{TypicalDualBij}
\begin{center}
\psset{yunit=0.32cm,xunit=0.28cm}
\begin{pspicture}(0,-0.5)(46,7)
%dashed grid
\psset{linewidth=0.25pt,linestyle=dashed, dash=2.5pt 1.5pt,linecolor=gray}
\psline{-}(0,2)(46,2) \psline{-}(0,3)(46,3) \psline{-}(0,4)(46,4)
\psline{-}(0,5)(46,5)
\psline{-}(1,1)(1,6) \psline{-}(2,1)(2,6) \psline{-}(3,1)(3,6)
\psline{-}(4,1)(4,6) \psline{-}(5,1)(5,6) \psline{-}(6,1)(6,6)
\psline{-}(7,1)(7,6) \psline{-}(8,1)(8,6) \psline{-}(9,1)(9,6)
\psline{-}(10,1)(10,6) \psline{-}(11,1)(11,6) \psline{-}(12,1)(12,6)
\psline{-}(13,1)(13,6) \psline{-}(14,1)(14,6) \psline{-}(15,1)(15,6)
\psline{-}(16,1)(16,6) \psline{-}(17,1)(17,6) \psline{-}(18,1)(18,6)
\psline{-}(19,1)(19,6) \psline{-}(20,1)(20,6) \psline{-}(21,1)(21,6)
\psline{-}(22,1)(22,6) \psline{-}(23,1)(23,6) \psline{-}(24,1)(24,6)
\psline{-}(25,1)(25,6) \psline{-}(26,1)(26,6) \psline{-}(27,1)(27,6)
\psline{-}(28,1)(28,6) \psline{-}(29,1)(29,6) \psline{-}(30,1)(30,6)
\psline{-}(31,1)(31,6) \psline{-}(32,1)(32,6) \psline{-}(33,1)(33,6)
\psline{-}(34,1)(34,6) \psline{-}(35,1)(35,6) \psline{-}(36,1)(36,6)
\psline{-}(37,1)(37,6) \psline{-}(38,1)(38,6) \psline{-}(39,1)(39,6)
\psline{-}(40,1)(40,6) \psline{-}(41,1)(41,6) \psline{-}(42,1)(42,6)
\psline{-}(43,1)(43,6) \psline{-}(44,1)(44,6) \psline{-}(45,1)(45,6)
\psline{-}(46,1)(46,6)
%axes
\psset{linewidth=0.25pt,fillstyle=none,linestyle=solid,linecolor=black}
\psline{->}(0,1)(46.5,1)
\psline{-}(0,6)(46,6)
\psline{-}(0,1)(0,6)
%numeros
\rput(-0.5,1){\scriptsize $1$}\rput(-0.5,3){\scriptsize $2$}
\rput(-0.5,5){\scriptsize $3$}
\rput(0,0.5){\scriptsize $0$} \rput(2,0.5){\scriptsize $1$}
\rput(4,0.5){\scriptsize $2$} \rput(6,0.5){\scriptsize $3$}
\rput(8,0.5){\scriptsize $4$} \rput(10,0.5){\scriptsize $5$}
\rput(12,0.5){\scriptsize $6$} \rput(14,0.5){\scriptsize $7$}
\rput(16,0.5){\scriptsize $8$} \rput(18,0.5){\scriptsize $9$}
\rput(20,0.5){\scriptsize $10$} \rput(22,0.5){\scriptsize $11$}
\rput(24,0.5){\scriptsize $12$} \rput(26,0.5){\scriptsize $13$}
\rput(28,0.5){\scriptsize $14$} \rput(30,0.5){\scriptsize $15$}
\rput(32,0.5){\scriptsize $16$} \rput(34,0.5){\scriptsize $17$}
\rput(36,0.5){\scriptsize $18$} \rput(38,0.5){\scriptsize $19$}
\rput(40,0.5){\scriptsize $20$} \rput(42,0.5){\scriptsize $21$}
\rput(44,0.5){\scriptsize $22$} \rput(46,0.5){\scriptsize $23$}
%path
\psset{linewidth=0.7pt,fillstyle=none,linestyle=solid,linecolor=black}
\psline(0,1)(1,2)(2,1)(3,2)(4,3)(5,4)(6,3)(7,2)(8,1)(9,2)(10,3)(11,4)
 (12,3)(13,4)(14,5)(15,4)(16,3)(17,2)(18,1)(19,2)(20,3)(21,4)(22,3)
 (23,2)(24,1)(25,2)(26,1)(27,2)(28,1)(29,2)(30,3)(31,4)(32,5)(33,6)
 (34,5)(35,6)(36,5)(37,4)(38,3)(39,4)(40,5)(41,6)(42,5)(43,6)(44,5)
 (45,6)(46,5)
\end{pspicture}
\end{center}
\end{figure}
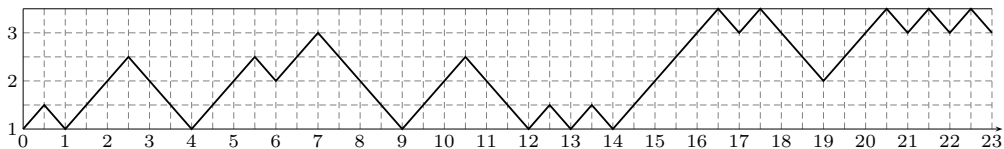

For the path $\hint$ in Fig.~\ref{TypicalDualInt},
each accretion vertex is highlighted using the symbol $\oplus$.
With $\nu=(11,8,4,4,2)$ as above, the insertion process then yields
the half-lattice path $\hh$ given in Fig.~\ref{TypicalDualBij}.
It may be checked that $\unwt(\hh)=229/2$ here.
Then, since $\unwt(\hhgs)=5/2$, we obtain $\wh(\hh)=112$.
As required, this equals the weight $\wt(h)$ of the RSOS path $h$
in Fig.~\ref{TypicalDualFig}.

That the map $h\mapsto\hh$ described above is a weight-preserving
bijection between the sets \eqref{Eq:Bijection2} is shown below.

\subsection{Proof of $\M(p,2p-1)$ bijection}  % Case 2

In what follows, we show that the combined map
\begin{equation}
\label{Eq:CombBij2}
h\to(\hcut,n,\lambda)
\to(\hlac,c,\mu,d,\nu)
\to(\hint,d,\nu)\to\hat h
\end{equation}
is a weight-preserving bijection between
$\PP^{p,2p-1}_{a,b-1}$ and $\HH^{p-1/2}_{\ha,\hb}$,
where we set $\ha=b/2$ and $\hb=a/2$.
To demonstrate the bijective property, it is sufficient to
demonstrate that the inverse of each step in \eqref{Eq:CombBij2}
is well-defined.

For the first step, $h$ is recovered from $(\hcut,n,\lambda)$ simply
by, for each $i\le n$, inserting a NE-SE or SE-NE pair of edges at the
scoring vertex $\lambda_i+1$ numbered amongst all the scoring vertices
of $\hcut$ counted from the right. The pair of edges is chosen so
that they lie in a dark band. If $\lambda_i$ is equal to the total
number of scoring vertices in $\hcut$, then a SE-NE edge pair is
inserted at the path startpoint $(0,a)$.

The inverse of the second step is achieved by truncating $\hlac$
at position $\hL(\hlac)$ reducing each peak in height by $1/2$ unit
by removing a NE-SE edge pair, then flipping the path horizontally,
and appending a $(b-1)$-tail.
The values of $m$ and $k$ may then be recovered from the resulting
$\hcut$, whereupon $\lambda$ may be recovered from $\mu$ and $\nu$
using \eqref{Eq:StaggerParts2} and \eqref{Eq:StaggerShift},
with $n=c+d$.

The characteristic property of $\hlac$ is that, to the left of
position $\hL(\hlac)$, every peak and valley is at integer height.
The characteristic property of $\hint$ is that, to the left of
position $\hL(\hint)$, every valley is at integer height,
while every peak, although not necessarily at integer height,
has its two neighbouring vertices both straight.
Therefore, for the inverse of the third step,
$(\hlac,c,\mu)$ is recovered from $\hint$ by reducing each
non-integer height peak by $1/2$ unit by removing a NE-SE edge
pair, and recording their positions $\mu_1,\mu_2,\ldots$
amongst the peaks of $\hint$, numbered from the left.

For the fourth step, note that the insertion of a NE-SE edge at an
accretion vertex of $\hint$ produces a new peak at non-integer height
for which the neighbouring two vertices are not both straight.
Since $\hint$ has no peaks of this nature, $\hint$ is thus recovered
from $\hh$ by removing the NE-SE edge pairs about each non-integer
height peak whose neighbouring vertices are not both straight.
The partition $\nu$ is then recovered by noting at which accretion
vertices of $\hint$, numbered from the right, these edge pairs have
been removed.

\subsection{Proof of $\M(p,2p-1)$ weight preservation (i)}
\label{Sec:Proof2_1}

The RSOS path $\hcut$ is obtained from $h$ by removing the
particles numbered $\lambda_1,\lambda_2,\ldots,\lambda_n$.
Consider the removal of the $i$th such particle from $h$.
The number of scoring vertices to its right is $\lambda_i$.
On removing the particle, the coordinates $(u_x,v_x)$ of
each scoring vertex to its right are reduced to $(u_x-1,v_x-1)$.
Therefore, from \eqref{Eq:RSOSWtDef}, we see that
this removal reduces the weight by $\lambda_i$.

Repeating this process for each of the $n$ particles in $h$ then
shows that
\begin{equation}
\label{Eq:RemFB2}
\wt(h^{\mathrm{cut}}) =\wt(h) - \sum_{i=1}^n \lambda_i.
\end{equation}

\subsection{Proof of $\M(p,2p-1)$ weight preservation (ii)}
\label{Sec:Proof2_2}

Obtain the half-lattice path $\hlac\in\HH^{p-1/2}_{\ha,\hb}$
from $\hcut\in\PP^{p,2p-1}_{a,b-1}$ as described above.
In this section, we show that $\wh(\hlac)=\wt(\hcut)$.

To do this, first regard the path $\hcut$, truncated after $L(\hcut)$
steps, as a sequence of lines alternating between
the SE and NE directions.
The $\M(p,2p-1)$ band structure, that $a$ and $b$ are even, and
the fact that the truncated $\hcut$ has no two neighbouring non-scoring
vertices implies that each line is of odd length except the first
if in the SE direction, and the last if in the NE direction.
The odd length lines are of length $2k'-1$ for $k'>0$ and
contain exactly $k'$ scoring vertices, if we include the final
vertex, and exclude the first vertex.
The possible even length lines are of length $2k'$ for $k'>0$
and also contain exactly $k'$ scoring vertices.
After flipping the path, and heightening the peaks to produce
the truncated $\hlac$, each of the odd length lines gives rise to
a line of length $k'$ which contains $2k'-1$ straight vertices,
and the possible even length lines give rise to a line of
length $k'$ which contains $2k'$ straight vertices
(including consideration of $(0,\ha)$).
Because of this, we may naturally associate each edge on a line of
the truncated $\hcut$ with a straight vertex on the corresponding
line of the truncated $\hlac$: associate the $i$th edge from the
left on the former with the $i$th straight vertex from the right
on the latter.

Through this association, we see that if an edge of $\hcut$ is
in the SE (resp.~NE) direction, then the number of up-scoring
(resp.~down-scoring) vertices to its right, is equal to
half the number of SE (resp.~NE) edges to
the left of the corresponding vertex in $\hlac$,
and to the right of $(0,\ha)$.
This will be used to prove that $\wh(\hlac)=\wt(\hcut)$,
once we have described alternative means of obtaining the
two weights.

In the case of a half-lattice path $\hh\in\HHtab$, for each vertex
$(x,\hh_{x})$, we define
\begin{equation}\label{Eq:HScoreWts}
\hu_x=\frac12 (x-\hh_x+\ha),\qquad
\hv_x=\frac12 (x+\hh_x-\ha),
\end{equation}
in analogy with \eqref{Eq:ScoreWts}.
Note that $\hu_x$ (resp.~$\hv_x$) is then half the number
of SE (resp.~NE) edges between $(0,\ha)$ and $(x,\hh_{x})$ in $\hh$.
Thus $\hu_x$ and $\hv_x$ are the numbers of up-scoring and
down-scoring vertices considered in the previous paragraph.

For any RSOS path $h$, define
\begin{equation}\label{Eq:EdgeWts}
w_x=
\begin{cases}
\text{number of up-scoring vertices to right of $(x,h_x)$}
&\text{if $h_x=h_{x-1}-1$,}\\ %SE
\text{number of down-scoring vertices to right of $(x,h_x)$}
&\text{if $h_x=h_{x-1}+1$,}   %NE
\end{cases}
\end{equation}
for all $x>0$.
Since the contribution of an up-scoring (resp.~down-scoring) vertex
to the weight $\wt(h)$ is given by $u_x$ (resp.~$v_x$) as defined
by \eqref{Eq:ScoreWts}, and this value is the number of SE (resp.~NE)
edges to its left, it follows that also
\begin{equation}\label{Eq:WtAlt}
\wt(h)=\sum_{x=1}^\infty w_x.
\end{equation}

As in Section \ref{Sec:Proof1_2},
we first consider the case where $a=b$,
and match each path segment of $\hcut$ with another
at the same height, one NE and one SE (the order is immaterial).
%as in Fig.~\ref{TypicalMatching}.
If they extend between vertices $(x-1,\hcut_x-1)$ and $(x,\hcut_x)$,
and vertices $(x'-1,\hcut_{x})$ and $(x',\hcut_{x}-1)$ respectively,
then in \eqref{Eq:WtAlt}, together they contribute $w_x+w_{x'}$
to $\wt(h)$.
Let, through the above association, the corresponding straight
vertices of $\hlac$ be $(\hx,\hlac_{\hx})$ and $(\hx',\hlac_{\hx})$
(they are at the same height),
and let their coordinates in the system \eqref{Eq:HScoreWts} be
$(\hu_{\hx},\hv_{\hx})$ and $(\hu'_{\hx'},\hv'_{\hx'})$ respectively.
Then the above association implies that
$w_x=\hv_{\hx}$ and $w_{x'}=\hu'_{\hx'}$, and therefore
\begin{equation}\label{Eq:FlipWts}
w_x+w_{x'}=\hv_{\hx}+\hu'_{\hx'}=\frac12(\hx+\hx').
\end{equation}
Since this is the contribution of these two straight vertices
to $\unwt(\hlac)$ using \eqref{Eq:WtsDef}, we conclude that
$\unwt(\hlac)=\wt(\hcut)$.
But $\wh(\hlac)=\unwt(\hlac)$ for $a=b$,
thereby proving that $\wh(\hlac)=\wt(\hcut)$ in this $a=b$ case.

In the $a\ne b$ case, proceed by extending the truncated $\hcut$
to the right
from $(L(\hcut),b)$ to $(L(\hcut)+|a-b|,a)$, so that this extended
path begins and ends at the same height. Then match the edges of
this extended path as before, but do not alter the values of $w_x$
obtained from \eqref{Eq:EdgeWts} for $\hcut$, where, in particular,
$w_x=0$ for $x>L(\hcut)$.

After flipping this extended path horizontally, shrinking by a factor
of 2, and heightening each peak by $1/2$ unit, we obtain the
truncated $\hlac$, but extended to the left of the vertical axis
by $|\ha-\hb|$ units,
as in the case of the path in Fig.~\ref{Fig:Extended}.

If the two matched edges from $\hcut$ are both to the left of
the extended portion, proceeding as in the $a=b$ case above
again yields \eqref{Eq:FlipWts}.
If one of the two matched edges is in the extended portion,
then the corresponding straight
vertex is in the extended portion of $\hlac$.
Then, if $x$, $x'$, $\hx$, $\hx'$, $(\hu_{\hx},\hv_{\hx})$ and
$(\hu'_{\hx'},\hv'_{\hx'})$
are obtained as above, either $x>L(\hcut)$ whereupon $w_x=0$ and
the definition \eqref{Eq:HScoreWts} gives $\hv_{\hx}=0$,
or $x'>L(\hcut)$ whereupon $w_{x'}=0$ and the definition
\eqref{Eq:HScoreWts} gives $\hu'_{\hx'}=0$.
Thus, \eqref{Eq:FlipWts} also holds in these cases.

Then, in view of \eqref{Eq:FlipWts},
we see that sum \eqref{Eq:WtAlt} for $\wt(\hcut)$
is equal to half the sum over the values of $\hx$
of the straight vertices of the extended half-lattice path $\hlac$.
Through the trick of Section \ref{Sec:WtTrick}, this sum gives
$\wh(\hlac)$, and therefore $\wh(\hlac)=\wt(\hcut)$
in this $a\ne b$ case.

\subsection{Proof of $\M(p,2p-1)$ weight preservation (iii)}
\label{Sec:Proof2_3}
On numbering the peaks of $\hlac$ from the left,
the half-lattice path $\hint$ is obtained from $\hlac$ by
raising each of the peaks numbered $\mu_1,\mu_2,\ldots,\mu_c$
by inserting a NE-SE edge pair. This is exactly the process
examined in Section \ref{Sec:Proof1_3}.
Therefore, from \eqref{Eq:DeeperVal}, we immediately obtain
\begin{equation}\label{Eq:DeeperVal2}
\wh(\hint)
=\wh(\hlac) +\frac{c}{2} (\numstr+c-1) + \sum_{i=1}^c \mu_i,
\end{equation}
where $\numstr$ is the number of straight vertices in $\hlac$.

\subsection{Proof of $\M(p,2p-1)$ weight preservation (iv)}
\label{Sec:Proof2_4}
Consider the insertion of a NE-SE edge pair at a vertex
of a half-lattice path which is either a valley or straight.
If it is a valley or straight-up, and there are $i$ straight vertices
to its right, the insertion moves each of these straight vertices
one unit to the right, resulting, via \eqref{Eq:WtsDef},
in a weight increase of $i/2$.
If it is a straight-down vertex, and there are $i-1$ straight vertices
to its right, the insertion moves each of these straight vertices
as well as the straight-down vertex itself one unit to the right,
thus also resulting in a weight increase of $i/2$.

With the accretion vertices of $\hint$ numbered from the right,
the specification of their positions ensures that the $j$th,
if a valley or a straight-up vertex, has exactly $2j$ straight vertices
to its right, and if straight-down, has exactly $2j-1$ straight
vertices to its right.

Thus, since the half-lattice path $\hh$ is obtained from $\hint$ by
inserting NE-SE edge pairs at the accretion vertices numbered
$\nu_1,\nu_2,\ldots,\nu_d$ (some of which may be equal),
we obtain
\begin{equation}\label{Eq:DeeperVal3}
\wh(\hh) =\wh(\hint) + \sum_{i=1}^d \nu_i.
\end{equation}

\subsection{Proof of $\M(p,2p-1)$ weight preservation (v)}
\label{Sec:Proof2_5}
We claim that $\numstr$, the number of straight vertices in $\hlac$,
is given by $\numstr=2k-2m$.
To see this, consider the association between the lines of
$\hcut$ and $\hlac$ described in the second paragraph of
Section \ref{Sec:Proof2_2}.

If the truncated $\hcut$ has $m$ peaks, let the lengths of
the lines to the left and right of the $i$th peak be
$2k^L_i-1$ and $2k^R_i-1$, for $1\le i\le m$.
Altogether, these lines contain
$k^L_1+\cdots+k^L_m+k^R_1+\cdots+k^R_m$ scoring vertices.
The corresponding lines in $\hlac$ then contain
$2(k^L_1+\cdots+k^L_m+k^R_1+\cdots+k^R_m)-2m$ straight vertices.
This proves the claim when the first and last lines
of the truncated $\hcut$ are in the NE and SE directions
respectively.

If the first line is in the SE direction and is of length $2k^R_0$,
then $\hcut$ has an additional $k^R_0$ scoring vertices,
and $\hlac$ has an additional $2k^R_0$ straight vertices
(including that at $(0,\ha)$).
The claim then follows in this case as well.

Likewise, if the last line of the truncated $\hcut$ is in
the NE direction and is of length $2k^L_0$,
then $\hcut$ has an additional $k^L_0$ scoring vertices,
and $\hlac$ has an additional $2k^L_0$ straight vertices.
It follows that the claim holds in all cases.

Using \eqref{Eq:RemFB2}, \eqref{Eq:DeeperVal2} and
\eqref{Eq:DeeperVal3}, and the
fact that $\wh(\hlac)=\wt(\hcut)$, we now obtain
\begin{equation}\label{Eq:WtDiff2}
\begin{split}
\wh(\hh)-\wt(h)
&=\sum_{i=1}^d \nu_i
+\frac c2(2k-2m+c-1)
+\sum_{i=1}^c \mu_i
-\sum_{i=1}^n \lambda_i\\
&=
c(k-m)+\frac c2(c-1)
+\sum_{i=1}^c \mu_i
-\sum_{i=1}^c \lambda_i\\
&=0,
\end{split}
\end{equation}
having used \eqref{Eq:StaggerShift}, and
where the final equality follows from \eqref{Eq:StaggerParts2} because
\begin{equation}
\sum_{i=1}^c\lambda_i-\sum_{i=1}^c\mu_i
=c(k-m)+\sum_{i=0}^{c-1} i=c(k-m)+\frac 12c(c-1).
\end{equation}
Thus the weight-preserving nature of the bijection
\eqref{Eq:CombBij2}
has been proved.

\section{Particles in half-lattice paths and fermionic expressions}

\subsection{Fermionic expressions}

Here, we derive fermionic expressions for the Virasoro characters
$\chi^{p,2p+1}_{r,s}$ and $\chi^{p,2p-1}_{r,s}$ for the
cases $r=1$ and $s=2$.

\begin{theorem}\label{Thm:HalfFerm}
For $t\in\HZZ$,
\begin{equation}
\chi^{t,2t+1}_{1,2}\equiv
\chi^{t+1/2,2t}_{1,2}=
\sum_{\boldn\in\ZZ_{\ge0}^{2t-3}}
  \frac{q^{\frac12{\boldn}\boldB^{(t)}{\boldn}^T}}{(q)_{m_{1}}}
  \prod_{j=2}^{2t-3}
  \qbinom{n_{j}+m_{j}}{n_{j}},
\end{equation}
where the sum is over vectors
$\boldn=(n_2,n_{3},n_4,\ldots,n_{2t-2})$,
$\boldB^{(t)}$ is the $(2t-3)\times(2t-3)$ symmetric matrix defined
by
\begin{equation}
\label{Eq:Bdef}
\boldB^{(t)}_{ij}=(i-1)j \qquad\text{for }i\le j,
\quad i,j\in\{2,3,4,\ldots,2t-2\}
\end{equation}
{\upshape(}$\boldB^{(t)}$ is the inverse of the Cartan matrix of
type $A_{2t-3}${\upshape)},
and
\begin{equation}
\label{Eq:Mdef}
m_d=\sum_{k=d+1}^{2t-2}
n_{k}(k-d),
\qquad\text{for } d\in\{1,2,3,\ldots, 2t-3\}.
\end{equation}
\end{theorem}

% Species t-1 doesn't move; there are an infinite number of charge 1/2.

\noindent
By virtue of Theorem \ref{Thm:HalfChar}, this may be proved by deriving
fermionic expressions for the generating functions $Y^t_{1,1}(q)$ for
half-lattice paths.
We accomplish this below by applying the method that was developed in
\cite{OleJS1} (extending that of \cite{Bres})
to obtain fermionic expressions for the ABF paths.
Its applicability relies on $\wh(\hh)$ being a multiple of
the sum of the $x$-positions of the straight vertices of $\hh$.

\subsection{Identifying particles}

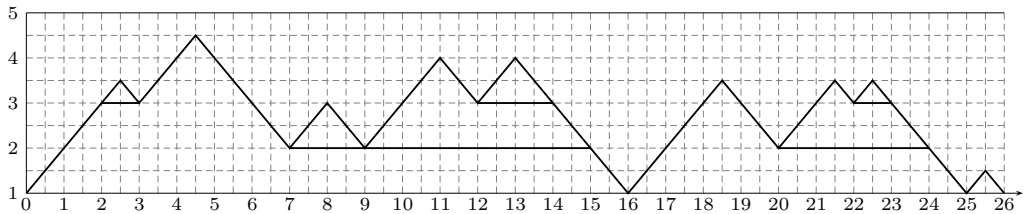
\begin{figure}[ht]
\caption{Dissecting a half-lattice path into particles}
\vskip-1mm
\label{Fig:Dissect}
\begin{center}
\psset{yunit=0.30cm,xunit=0.25cm}
\begin{pspicture}(0,-0.5)(52,10)
%dashed grid
\psset{linewidth=0.25pt,linestyle=dashed, dash=2.5pt 1.5pt,linecolor=gray}
\psline{-}(0,2)(52,2) \psline{-}(0,3)(52,3) \psline{-}(0,4)(52,4)
\psline{-}(0,5)(52,5) \psline{-}(0,6)(52,6) \psline{-}(0,7)(52,7)
\psline{-}(0,8)(52,8)
\psline{-}(1,1)(1,9) \psline{-}(2,1)(2,9) \psline{-}(3,1)(3,9)
\psline{-}(4,1)(4,9) \psline{-}(5,1)(5,9) \psline{-}(6,1)(6,9)
\psline{-}(7,1)(7,9) \psline{-}(8,1)(8,9) \psline{-}(9,1)(9,9)
\psline{-}(10,1)(10,9) \psline{-}(11,1)(11,9) \psline{-}(12,1)(12,9)
\psline{-}(13,1)(13,9) \psline{-}(14,1)(14,9) \psline{-}(15,1)(15,9)
\psline{-}(16,1)(16,9) \psline{-}(17,1)(17,9) \psline{-}(18,1)(18,9)
\psline{-}(19,1)(19,9) \psline{-}(20,1)(20,9) \psline{-}(21,1)(21,9)
\psline{-}(22,1)(22,9) \psline{-}(23,1)(23,9) \psline{-}(24,1)(24,9)
\psline{-}(25,1)(25,9) \psline{-}(26,1)(26,9) \psline{-}(27,1)(27,9)
\psline{-}(28,1)(28,9) \psline{-}(29,1)(29,9) \psline{-}(30,1)(30,9)
\psline{-}(31,1)(31,9) \psline{-}(32,1)(32,9) \psline{-}(33,1)(33,9)
\psline{-}(34,1)(34,9) \psline{-}(35,1)(35,9) \psline{-}(36,1)(36,9)
\psline{-}(37,1)(37,9) \psline{-}(38,1)(38,9) \psline{-}(39,1)(39,9)
\psline{-}(40,1)(40,9) \psline{-}(41,1)(41,9) \psline{-}(42,1)(42,9)
\psline{-}(43,1)(43,9) \psline{-}(44,1)(44,9) \psline{-}(45,1)(45,9)
\psline{-}(46,1)(46,9) \psline{-}(47,1)(47,9) \psline{-}(48,1)(48,9)
\psline{-}(49,1)(49,9) \psline{-}(50,1)(50,9) \psline{-}(51,1)(51,9)
\psline{-}(52,1)(52,9)
%axes
\psset{linewidth=0.25pt,fillstyle=none,linestyle=solid,linecolor=black}
\psline{->}(0,1)(53,1)
\psline{-}(0,9)(52,9)
\psline{-}(0,1)(0,9)
%numeros
\rput(-0.7,1){\scriptsize $1$}\rput(-0.7,3){\scriptsize $2$}
\rput(-0.7,5){\scriptsize $3$} \rput(-0.7,7){\scriptsize $4$}
\rput(-0.7,9){\scriptsize $5$}
\rput(0,0.5){\scriptsize $0$} \rput(2,0.5){\scriptsize $1$}
\rput(4,0.5){\scriptsize $2$} \rput(6,0.5){\scriptsize $3$}
\rput(8,0.5){\scriptsize $4$} \rput(10,0.5){\scriptsize $5$}
\rput(12,0.5){\scriptsize $6$} \rput(14,0.5){\scriptsize $7$}
\rput(16,0.5){\scriptsize $8$} \rput(18,0.5){\scriptsize $9$}
\rput(20,0.5){\scriptsize $10$} \rput(22,0.5){\scriptsize $11$}
\rput(24,0.5){\scriptsize $12$} \rput(26,0.5){\scriptsize $13$}
\rput(28,0.5){\scriptsize $14$} \rput(30,0.5){\scriptsize $15$}
\rput(32,0.5){\scriptsize $16$} \rput(34,0.5){\scriptsize $17$}
\rput(36,0.5){\scriptsize $18$} \rput(38,0.5){\scriptsize $19$}
\rput(40,0.5){\scriptsize $20$} \rput(42,0.5){\scriptsize $21$}
\rput(44,0.5){\scriptsize $22$} \rput(46,0.5){\scriptsize $23$}
\rput(48,0.5){\scriptsize $24$} \rput(50,0.5){\scriptsize $25$}
\rput(52,0.5){\scriptsize $26$}
%path
\psset{linewidth=0.7pt,fillstyle=none,linestyle=solid,linecolor=black}
\psline(0,1)(5,6)(6,5)(9,8)(14,3)(16,5)(18,3)
(22,7)(24,5)(26,7)(32,1)
(37,6)(40,3)(43,6)(44,5)(45,6)(50,1)(51,2)(52,1)
%baseline
\psset{linewidth=0.7pt,fillstyle=none,linestyle=solid,linecolor=black}
\psline(14,3)(30,3)
\psline(40,3)(48,3)
\psline(4,5)(6,5)
\psline(24,5)(28,5)
\psline(44,5)(46,5)
%dots
\psset{dotsize=2.5pt,fillcolor=black}
%\psdots(0,3)(1,2)(5,2)(6,3)(7,4)(9,4)(11,4)(12,5)(13,6)(14,7)
%(17,8)(19,8)(22,7)(23,6)(27,6)(28,7)(29,8)(31,8)
%gs
\psset{linewidth=0.7pt,linestyle=dashed, dash=4pt 2pt,linecolor=black}
%\psline(0,3)(1,4)(2,5)(3,6)(4,7)(5,8)(6,7)(7,8)(8,7)(9,8)(10,7)(11,8)(12,7)
\psset{dotsize=2.5pt,dotstyle=o,fillcolor=white}
%\psdots(1,4)(2,5)(3,6)(4,7)
\end{pspicture}
\end{center}
\end{figure}

To prove Theorem \ref{Thm:HalfFerm}, we first dissect each
half-lattice path $\hh\in\HH^t_{1,1}$ into particles
of various charges as follows
(these particles are not related to the particles described in the
bijections of Sections \ref{Sec:Bijection1} and \ref{Sec:Bijection2}).
We assign each peak of $\hh$ a charge $\hat d\in\{1/2,1,3/2,2,\ldots,t-1\}$,
and associate a particle of that charge with the peak.
This is done recursively starting with the peaks of charge 1/2.
Scanning from the right, identify the vertical distances
from the peak to the two neighbouring valleys.
The peak is assigned a charge 1/2 if either of those distances is 1/2.
To delineate the particle, a horizontal \emph{baseline} of length 1
is then drawn symmetrically about the peak at a height $1/2$ less
than the peak.
This baseline will intersect the path twice, with at least one
intersection at a valley.
In proceeding from the right, a single valley on a baseline
is discounted from further consideration
(if two, discount only one).
Thus, in Fig.~\ref{Fig:Dissect} for example, having assigned
the peak $(45/2,7/2)$ a charge $1/2$,
and drawn its baseline, the valley $(22,4)$ cannot be used
to identify the peak $(43/2,7/2)$ as charge $1/2$.
In this case, the neighbouring valleys are now $(20,2)$ and $(25,1)$.

Because of the infinite tail, there will be an infinite number of
peaks, and thus particles, of charge 1/2.
The remaining finite number of peaks are then identified
with particles of greater charge.
Having identified all the peaks having charge $1/2$, we then
scan from the right for peaks having vertical distance 1 to
one of their neighbouring two valleys, ignoring those valleys
discounted previously.
For each such peak,
we draw a horizontal baseline at a height 1 less than that of the peak,
below the peak, extending from the identified valley to
where it next intersects the path.
These identify the particles of charge 1.
Then, continuing in this fashion, next considering charges
$\hat d=3/2,2,5/2,\ldots,t-1$, in turn, we eventually assign a
charge and a particle to each peak, with its baseline of
length $2\hat d$ at height $\hat d$ below that of the peak.
We define the \emph{origin} of the particle to be the position
of the left end of its baseline.
Note that the baselines are necessarily all at integer heights.
The slope of a particle is defined to be that portion of the path
that lies above its baseline, but not above other baselines.

This dissection process has been carried out for the path
given in Fig.~\ref{Fig:Dissect}.
From left to right, its peaks have been assigned charges
$\tfrac12,\tfrac72,1,2,1,\tfrac52,
\tfrac32,\tfrac12,\tfrac12,
 \tfrac12,\ldots$.

Although enacted differently to 
the dissection process described in \cite[Section 3.1]{OleJS1},
the two processes are equivalent.

\begin{figure}[ht]
\caption{Minimal weight path $\hh^{(2,1,1,1,0,1,0)}\in\HH^5_{1,1}$}
\vskip-1mm
\label{Fig:MinimalConf}
\begin{center}
\psset{yunit=0.30cm,xunit=0.25cm}
\begin{pspicture}(0,-0.5)(52,10)
%dashed grid
\psset{linewidth=0.25pt,linestyle=dashed, dash=2.5pt 1.5pt,linecolor=gray}
\psline{-}(0,2)(52,2) \psline{-}(0,3)(52,3) \psline{-}(0,4)(52,4)
\psline{-}(0,5)(52,5) \psline{-}(0,6)(52,6) \psline{-}(0,7)(52,7)
\psline{-}(0,8)(52,8)
\psline{-}(1,1)(1,9) \psline{-}(2,1)(2,9) \psline{-}(3,1)(3,9)
\psline{-}(4,1)(4,9) \psline{-}(5,1)(5,9) \psline{-}(6,1)(6,9)
\psline{-}(7,1)(7,9) \psline{-}(8,1)(8,9) \psline{-}(9,1)(9,9)
\psline{-}(10,1)(10,9) \psline{-}(11,1)(11,9) \psline{-}(12,1)(12,9)
\psline{-}(13,1)(13,9) \psline{-}(14,1)(14,9) \psline{-}(15,1)(15,9)
\psline{-}(16,1)(16,9) \psline{-}(17,1)(17,9) \psline{-}(18,1)(18,9)
\psline{-}(19,1)(19,9) \psline{-}(20,1)(20,9) \psline{-}(21,1)(21,9)
\psline{-}(22,1)(22,9) \psline{-}(23,1)(23,9) \psline{-}(24,1)(24,9)
\psline{-}(25,1)(25,9) \psline{-}(26,1)(26,9) \psline{-}(27,1)(27,9)
\psline{-}(28,1)(28,9) \psline{-}(29,1)(29,9) \psline{-}(30,1)(30,9)
\psline{-}(31,1)(31,9) \psline{-}(32,1)(32,9) \psline{-}(33,1)(33,9)
\psline{-}(34,1)(34,9) \psline{-}(35,1)(35,9) \psline{-}(36,1)(36,9)
\psline{-}(37,1)(37,9) \psline{-}(38,1)(38,9) \psline{-}(39,1)(39,9)
\psline{-}(40,1)(40,9) \psline{-}(41,1)(41,9) \psline{-}(42,1)(42,9)
\psline{-}(43,1)(43,9) \psline{-}(44,1)(44,9) \psline{-}(45,1)(45,9)
\psline{-}(46,1)(46,9) \psline{-}(47,1)(47,9) \psline{-}(48,1)(48,9)
\psline{-}(49,1)(49,9) \psline{-}(50,1)(50,9) \psline{-}(51,1)(51,9)
\psline{-}(52,1)(52,9)
%axes
\psset{linewidth=0.25pt,fillstyle=none,linestyle=solid,linecolor=black}
\psline{->}(0,1)(53,1)
\psline{-}(0,9)(52,9)
\psline{-}(0,1)(0,9)
%numeros
\rput(-0.7,1){\scriptsize $1$}\rput(-0.7,3){\scriptsize $2$}
\rput(-0.7,5){\scriptsize $3$} \rput(-0.7,7){\scriptsize $4$}
\rput(-0.7,9){\scriptsize $5$}
\rput(0,0.5){\scriptsize $0$} \rput(2,0.5){\scriptsize $1$}
\rput(4,0.5){\scriptsize $2$} \rput(6,0.5){\scriptsize $3$}
\rput(8,0.5){\scriptsize $4$} \rput(10,0.5){\scriptsize $5$}
\rput(12,0.5){\scriptsize $6$} \rput(14,0.5){\scriptsize $7$}
\rput(16,0.5){\scriptsize $8$} \rput(18,0.5){\scriptsize $9$}
\rput(20,0.5){\scriptsize $10$} \rput(22,0.5){\scriptsize $11$}
\rput(24,0.5){\scriptsize $12$} \rput(26,0.5){\scriptsize $13$}
\rput(28,0.5){\scriptsize $14$} \rput(30,0.5){\scriptsize $15$}
\rput(32,0.5){\scriptsize $16$} \rput(34,0.5){\scriptsize $17$}
\rput(36,0.5){\scriptsize $18$} \rput(38,0.5){\scriptsize $19$}
\rput(40,0.5){\scriptsize $20$} \rput(42,0.5){\scriptsize $21$}
\rput(44,0.5){\scriptsize $22$} \rput(46,0.5){\scriptsize $23$}
\rput(48,0.5){\scriptsize $24$} \rput(50,0.5){\scriptsize $25$}
\rput(52,0.5){\scriptsize $26$}
%path
\psset{linewidth=0.7pt,fillstyle=none,linestyle=solid,linecolor=black}
\psline(0,1)(7,8)(14,1)(19,6)(24,1)(28,5)(32,1)
(35,4)(38,1)(40,3)(42,1)(44,3)(46,1)
(47,2)(48,1)(49,2)(50,1)(51,2)(52,1)
%dots
\psset{dotsize=2.5pt,fillcolor=black}
%\psdots(0,3)(1,2)(5,2)(6,3)(7,4)(9,4)(11,4)(12,5)(13,6)(14,7)
%(17,8)(19,8)(22,7)(23,6)(27,6)(28,7)(29,8)(31,8)
%gs
\psset{linewidth=0.7pt,linestyle=dashed, dash=4pt 2pt,linecolor=black}
%\psline(0,3)(1,4)(2,5)(3,6)(4,7)(5,8)(6,7)(7,8)(8,7)(9,8)(10,7)(11,8)(12,7)
\psset{dotsize=2.5pt,dotstyle=o,fillcolor=white}
%\psdots(1,4)(2,5)(3,6)(4,7)
\end{pspicture}
\end{center}
\end{figure}
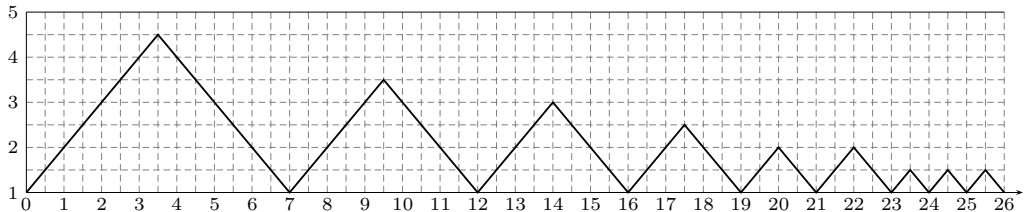

For $\boldn=(n_2,n_{3},n_4,\ldots,n_{2t-2})$,
define $\sector{\boldn}\subset\HH^t_{1,1}$ to be the set of
half-lattice paths that contain $n_d$ particles of charge $d/2$,
%for $c=1,3/2,2,\ldots,t-1$.
for $2\le d\le 2t-2$.
Then define $\hh^{\boldn}\in\sector{\boldn}$ to be that element
that has minimal weight amongst all the elements of $\sector{\boldn}$.
By considering the possible positions of the straight vertices
in $\hh^{\boldn}$, it can be seen that the minimal weight
condition implies that $\hh^{\boldn}$ is unique, that the
baseline of each of the particles in $\hh^{\boldn}$ lies at height 1,
and each particle is to the left of all particles of lesser charge.
For example, for $\boldn=(2,1,1,1,0,1,0)$, the minimal weight path
$\hh^{\boldn}\in\HH^5_{1,1}$ is given in Fig.~\ref{Fig:MinimalConf}.

\begin{lemma}\label{Lem:MinimalWt}
Let $t\in\frac12\ZZ_{\ge4}$ and
$\boldn=(n_2,n_{3},n_4,\ldots,n_{2t-2})$.
Then, with $\boldB^{(t)}$ given by \eqref{Eq:Bdef},
\begin{equation}
\label{Eq:MinimalWt}
\wh(\hh^{\boldn}) = \frac12{\boldn}\boldB^{(t)}{\boldn}^T.
\end{equation}
\end{lemma}

\Proof First note that if a particle of charge $d/2$ has its
origin at $(x,h_x)$, then its $2(d-1)$ straight vertices contribute
\begin{equation}
\begin{split}
\tfrac12\big(
&(x+\tfrac12)+(x+1)+(x+\tfrac32)+\cdots+(x+\tfrac d2-\tfrac12)
+(x+\tfrac d2+\tfrac12)+(x+\tfrac d2+1)\\
&\qquad\qquad+
(x+\tfrac d2+\tfrac32)+\cdots+(x+d-\tfrac12)
\big)
=(x+\tfrac d2)(d-1)
\end{split}
\end{equation}
to $\wh(\hh^{\boldn})$.
Then, a succession of $n_d$ particles of charge $d/2$, with the
origin of the first at $(x,h_x)$, contribute
\begin{equation}
\label{Eq:SingleChargeWt}
\sum_{i=1}^{n_d} (x+(i-\tfrac12)d)(d-1)
=n_d(x+\tfrac12dn_d)(d-1)
\end{equation}
to $\wh(\hh^{\boldn})$.
Now note that the total length of the baselines
of the particles of charges $t-1,t-\tfrac32,t-2,\ldots,\tfrac12(d+1)$
is $x=\sum_{j=d+1}^{2t-2}jn_{j}$.
After substituting this into \eqref{Eq:SingleChargeWt}, and summing
over $d=2,3,4,\ldots,2t-2$,
expression \eqref{Eq:MinimalWt} results.
\cqfd

\subsection{Particle moves}
Here we describe a process in which the particles in a path $\hh$
may move.
As we will see below, all paths in sector $\sector{\boldn}$
may be generated from the minimal weight path $\hh^{\boldn}$,
through this process.

A particle of charge $\hat d$ has a permitted move if its
baseline has length exactly $2\hat d$, and its origin
%is not at $x=0$ and
intersects the slope of a particle of greater charge $\hat d'>\hat d$.
Let the peaks of these two particles be at $(x,\hh_x)$
and $(x',\hh_{x'})$ respectively.
Note that the restriction that valleys are at integer height
implies that $\hh_{x'}-\hh_{x}$ is an integer if and only if
$\hat d'-\hat d$ is an integer.

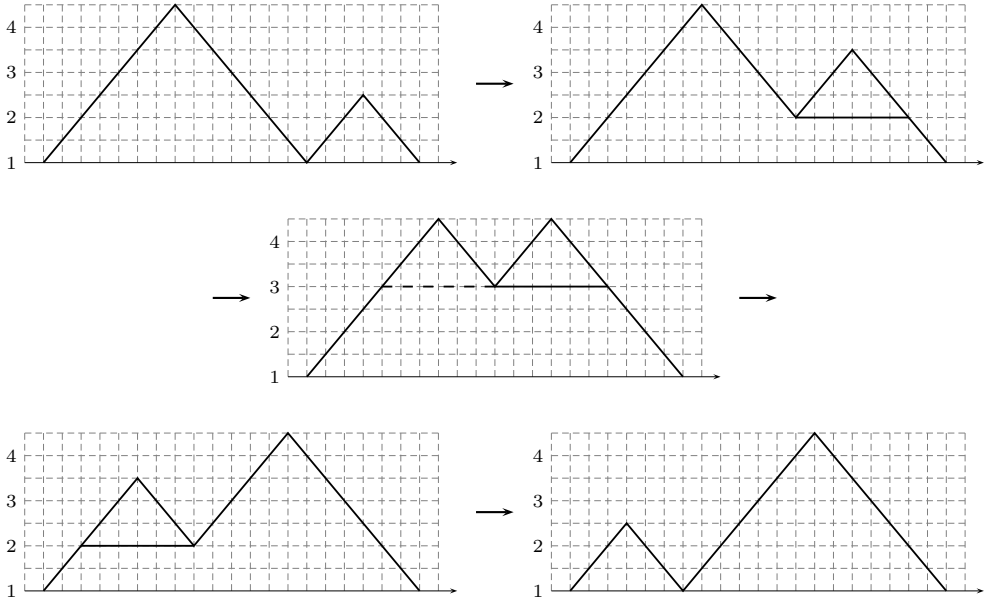
\begin{figure}[ht]
\caption{Move sequence for $c=3/2$, $c'=7/2$.}
\vskip-1mm
\label{Fig:Moves1}
\begin{center}
\psset{yunit=0.30cm,xunit=0.25cm}
\begin{pspicture}(-1,-19.5)(50,9)
%dashed grid
\rput[bl](0,0){
\psset{linewidth=0.25pt,linestyle=dashed, dash=2.5pt 1.5pt,linecolor=gray}
\psline{-}(0,2)(22,2) \psline{-}(0,3)(22,3) \psline{-}(0,4)(22,4)
\psline{-}(0,5)(22,5) \psline{-}(0,6)(22,6) \psline{-}(0,7)(22,7)
\psline{-}(0,8)(22,8)
\psline{-}(0,1)(0,8)
\psline{-}(1,1)(1,8) \psline{-}(2,1)(2,8) \psline{-}(3,1)(3,8)
\psline{-}(4,1)(4,8) \psline{-}(5,1)(5,8) \psline{-}(6,1)(6,8)
\psline{-}(7,1)(7,8) \psline{-}(8,1)(8,8) \psline{-}(9,1)(9,8)
\psline{-}(10,1)(10,8) \psline{-}(11,1)(11,8) \psline{-}(12,1)(12,8)
\psline{-}(13,1)(13,8) \psline{-}(14,1)(14,8) \psline{-}(15,1)(15,8)
\psline{-}(16,1)(16,8) \psline{-}(17,1)(17,8) \psline{-}(18,1)(18,8)
\psline{-}(19,1)(19,8) \psline{-}(20,1)(20,8) \psline{-}(21,1)(21,8)
\psline{-}(22,1)(22,8)
%axes
\psset{linewidth=0.25pt,fillstyle=none,linestyle=solid,linecolor=black}
\psline{->}(0,1)(23,1)
%\psline{-}(0,9)(22,9)
%numeros
\rput(-0.7,1){\scriptsize $1$}\rput(-0.7,3){\scriptsize $2$}
\rput(-0.7,5){\scriptsize $3$} \rput(-0.7,7){\scriptsize $4$}
%path
\psset{linewidth=0.7pt,fillstyle=none,linestyle=solid,linecolor=black}
\psline(1,1)(8,8)(15,1)(18,4)(21,1)
%dots
\psset{dotsize=2.5pt,fillcolor=black}
}
\psline{->}(24,4.5)(26,4.5)
\rput[bl](28,0){
\psset{linewidth=0.25pt,linestyle=dashed, dash=2.5pt 1.5pt,linecolor=gray}
\psline{-}(0,2)(22,2) \psline{-}(0,3)(22,3) \psline{-}(0,4)(22,4)
\psline{-}(0,5)(22,5) \psline{-}(0,6)(22,6) \psline{-}(0,7)(22,7)
\psline{-}(0,8)(22,8)
\psline{-}(0,1)(0,8)
\psline{-}(1,1)(1,8) \psline{-}(2,1)(2,8) \psline{-}(3,1)(3,8)
\psline{-}(4,1)(4,8) \psline{-}(5,1)(5,8) \psline{-}(6,1)(6,8)
\psline{-}(7,1)(7,8) \psline{-}(8,1)(8,8) \psline{-}(9,1)(9,8)
\psline{-}(10,1)(10,8) \psline{-}(11,1)(11,8) \psline{-}(12,1)(12,8)
\psline{-}(13,1)(13,8) \psline{-}(14,1)(14,8) \psline{-}(15,1)(15,8)
\psline{-}(16,1)(16,8) \psline{-}(17,1)(17,8) \psline{-}(18,1)(18,8)
\psline{-}(19,1)(19,8) \psline{-}(20,1)(20,8) \psline{-}(21,1)(21,8)
\psline{-}(22,1)(22,8)
%axes
\psset{linewidth=0.25pt,fillstyle=none,linestyle=solid,linecolor=black}
\psline{->}(0,1)(23,1)
%\psline{-}(0,9)(22,9)
%numeros
\rput(-0.7,1){\scriptsize $1$}\rput(-0.7,3){\scriptsize $2$}
\rput(-0.7,5){\scriptsize $3$} \rput(-0.7,7){\scriptsize $4$}
%path
\psset{linewidth=0.7pt,fillstyle=none,linestyle=solid,linecolor=black}
\psline(1,1)(8,8)(13,3)(16,6)(21,1)
\psline(13,3)(19,3)
%dots
\psset{dotsize=2.5pt,fillcolor=black}
}
\psline{->}(10,-5)(12,-5)
\rput[bl](14,-9.5){
\psset{linewidth=0.25pt,linestyle=dashed, dash=2.5pt 1.5pt,linecolor=gray}
\psline{-}(0,2)(22,2) \psline{-}(0,3)(22,3) \psline{-}(0,4)(22,4)
\psline{-}(0,5)(22,5) \psline{-}(0,6)(22,6) \psline{-}(0,7)(22,7)
\psline{-}(0,8)(22,8)
\psline{-}(0,1)(0,8)
\psline{-}(1,1)(1,8) \psline{-}(2,1)(2,8) \psline{-}(3,1)(3,8)
\psline{-}(4,1)(4,8) \psline{-}(5,1)(5,8) \psline{-}(6,1)(6,8)
\psline{-}(7,1)(7,8) \psline{-}(8,1)(8,8) \psline{-}(9,1)(9,8)
\psline{-}(10,1)(10,8) \psline{-}(11,1)(11,8) \psline{-}(12,1)(12,8)
\psline{-}(13,1)(13,8) \psline{-}(14,1)(14,8) \psline{-}(15,1)(15,8)
\psline{-}(16,1)(16,8) \psline{-}(17,1)(17,8) \psline{-}(18,1)(18,8)
\psline{-}(19,1)(19,8) \psline{-}(20,1)(20,8) \psline{-}(21,1)(21,8)
\psline{-}(22,1)(22,8)
%axes
\psset{linewidth=0.25pt,fillstyle=none,linestyle=solid,linecolor=black}
\psline{->}(0,1)(23,1)
%\psline{-}(0,9)(22,9)
%numeros
\rput(-0.7,1){\scriptsize $1$}\rput(-0.7,3){\scriptsize $2$}
\rput(-0.7,5){\scriptsize $3$} \rput(-0.7,7){\scriptsize $4$}
%path
\psset{linewidth=0.7pt,fillstyle=none,linestyle=solid,linecolor=black}
\psline(1,1)(8,8)(11,5)(14,8)(21,1)
\psline(11,5)(17,5)
\psset{linewidth=0.7pt,fillstyle=none,linestyle=dashed,dash=4pt 4pt,
linecolor=black}
\psline(5,5)(11,5)
%dots
\psset{dotsize=2.5pt,fillcolor=black}
}
\psline{->}(38,-5)(40,-5)
\rput[bl](0,-19){
\psset{linewidth=0.25pt,linestyle=dashed, dash=2.5pt 1.5pt,linecolor=gray}
\psline{-}(0,2)(22,2) \psline{-}(0,3)(22,3) \psline{-}(0,4)(22,4)
\psline{-}(0,5)(22,5) \psline{-}(0,6)(22,6) \psline{-}(0,7)(22,7)
\psline{-}(0,8)(22,8)
\psline{-}(0,1)(0,8)
\psline{-}(1,1)(1,8) \psline{-}(2,1)(2,8) \psline{-}(3,1)(3,8)
\psline{-}(4,1)(4,8) \psline{-}(5,1)(5,8) \psline{-}(6,1)(6,8)
\psline{-}(7,1)(7,8) \psline{-}(8,1)(8,8) \psline{-}(9,1)(9,8)
\psline{-}(10,1)(10,8) \psline{-}(11,1)(11,8) \psline{-}(12,1)(12,8)
\psline{-}(13,1)(13,8) \psline{-}(14,1)(14,8) \psline{-}(15,1)(15,8)
\psline{-}(16,1)(16,8) \psline{-}(17,1)(17,8) \psline{-}(18,1)(18,8)
\psline{-}(19,1)(19,8) \psline{-}(20,1)(20,8) \psline{-}(21,1)(21,8)
\psline{-}(22,1)(22,8)
%axes
\psset{linewidth=0.25pt,fillstyle=none,linestyle=solid,linecolor=black}
\psline{->}(0,1)(23,1)
%\psline{-}(0,9)(22,9)
%numeros
\rput(-0.7,1){\scriptsize $1$}\rput(-0.7,3){\scriptsize $2$}
\rput(-0.7,5){\scriptsize $3$} \rput(-0.7,7){\scriptsize $4$}
%path
\psset{linewidth=0.7pt,fillstyle=none,linestyle=solid,linecolor=black}
\psline(1,1)(6,6)(9,3)(14,8)(21,1)
\psline(3,3)(9,3)
%dots
\psset{dotsize=2.5pt,fillcolor=black}
%\psdots(0,3)(1,2)(5,2)(6,3)(7,4)(9,4)(11,4)(12,5)(13,6)(14,7)
%(17,8)(19,8)(22,7)(23,6)(27,6)(28,7)(29,8)(31,8)
}
\psline{->}(24,-14.5)(26,-14.5)
\rput[bl](28,-19){
\psset{linewidth=0.25pt,linestyle=dashed, dash=2.5pt 1.5pt,linecolor=gray}
\psline{-}(0,2)(22,2) \psline{-}(0,3)(22,3) \psline{-}(0,4)(22,4)
\psline{-}(0,5)(22,5) \psline{-}(0,6)(22,6) \psline{-}(0,7)(22,7)
\psline{-}(0,8)(22,8)
\psline{-}(0,1)(0,8)
\psline{-}(1,1)(1,8) \psline{-}(2,1)(2,8) \psline{-}(3,1)(3,8)
\psline{-}(4,1)(4,8) \psline{-}(5,1)(5,8) \psline{-}(6,1)(6,8)
\psline{-}(7,1)(7,8) \psline{-}(8,1)(8,8) \psline{-}(9,1)(9,8)
\psline{-}(10,1)(10,8) \psline{-}(11,1)(11,8) \psline{-}(12,1)(12,8)
\psline{-}(13,1)(13,8) \psline{-}(14,1)(14,8) \psline{-}(15,1)(15,8)
\psline{-}(16,1)(16,8) \psline{-}(17,1)(17,8) \psline{-}(18,1)(18,8)
\psline{-}(19,1)(19,8) \psline{-}(20,1)(20,8) \psline{-}(21,1)(21,8)
\psline{-}(22,1)(22,8)
%axes
\psset{linewidth=0.25pt,fillstyle=none,linestyle=solid,linecolor=black}
\psline{->}(0,1)(23,1)
%\psline{-}(0,9)(22,9)
%numeros
\rput(-0.7,1){\scriptsize $1$}\rput(-0.7,3){\scriptsize $2$}
\rput(-0.7,5){\scriptsize $3$} \rput(-0.7,7){\scriptsize $4$}
%path
\psset{linewidth=0.7pt,fillstyle=none,linestyle=solid,linecolor=black}
\psline(1,1)(4,4)(7,1)(14,8)(21,1)
%dots
\psset{dotsize=2.5pt,fillcolor=black}
}
\end{pspicture}
\end{center}
\end{figure}

If $x<x'$ or $\hh_x\le\hh_{x'}-1$
then the move is performed by removing the two edges immediately
to the left of the particle of charge $\hat d$ 
(these are necessarily in the same direction),
then moving the particle itself two half-units to the left,
and two half-units either up or down, to fill the gap,
before finally, reinserting the originally removed two edges
at the right of the particle to recover a connected path.
This process has been carried out for the first, second and fourth
moves illustrated in Fig.~\ref{Fig:Moves1}.

If $x>x'$ and $\hh_x=\hh_{x'}$ then this same procedure is carried out
after, contrary to the above convention,
considering the peak at $(x',\hh_{x'})$
to be that of the particle of charge $\hat d$,
and the peak at $(x,\hh_{x})$ that of the particle of change $\hat d'$.
The baseline of the former is moved accordingly.
For example, in the third path picture of Fig.~\ref{Fig:Moves1},
the original definition associates particles
of charges 7/2 and 3/2 with the two peaks.
The baseline of the latter is shown as a solid line.
Before enacting its move, we first reassign the first peak as that
having a charge 3/2, with the dashed line as its baseline.
The move then results in the fourth path.

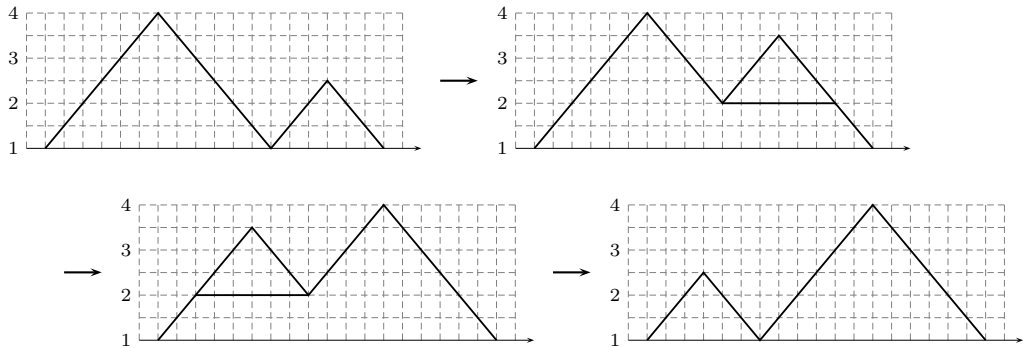
\begin{figure}[ht]
\caption{Move sequence for $\hat d=3/2$, $\hat d'=3$.}
\vskip-1mm
\label{Fig:Moves2}
\begin{center}
\psset{yunit=0.30cm,xunit=0.25cm}
\begin{pspicture}(-1,-9)(48,9)
\rput[bl](0,0){
%dashed grid
\psset{linewidth=0.25pt,linestyle=dashed, dash=2.5pt 1.5pt,linecolor=gray}
\psline{-}(0,2)(20,2) \psline{-}(0,3)(20,3) \psline{-}(0,4)(20,4)
\psline{-}(0,5)(20,5) \psline{-}(0,6)(20,6) \psline{-}(0,7)(20,7)
\psline{-}(0,1)(0,7)
\psline{-}(1,1)(1,7) \psline{-}(2,1)(2,7) \psline{-}(3,1)(3,7)
\psline{-}(4,1)(4,7) \psline{-}(5,1)(5,7) \psline{-}(6,1)(6,7)
\psline{-}(7,1)(7,7) \psline{-}(8,1)(8,7) \psline{-}(9,1)(9,7)
\psline{-}(10,1)(10,7) \psline{-}(11,1)(11,7) \psline{-}(12,1)(12,7)
\psline{-}(13,1)(13,7) \psline{-}(14,1)(14,7) \psline{-}(15,1)(15,7)
\psline{-}(16,1)(16,7) \psline{-}(17,1)(17,7) \psline{-}(18,1)(18,7)
\psline{-}(19,1)(19,7) \psline{-}(20,1)(20,7)
%axes
\psset{linewidth=0.25pt,fillstyle=none,linestyle=solid,linecolor=black}
\psline{->}(0,1)(21,1)
%\psline{-}(0,9)(22,9)
%numeros
\rput(-0.7,1){\scriptsize $1$}\rput(-0.7,3){\scriptsize $2$}
\rput(-0.7,5){\scriptsize $3$} \rput(-0.7,7){\scriptsize $4$}
%path
\psset{linewidth=0.7pt,fillstyle=none,linestyle=solid,linecolor=black}
\psline(1,1)(7,7)(13,1)(16,4)(19,1)
%dots
\psset{dotsize=2.5pt,fillcolor=black}
}
\psline{->}(22,4)(24,4)
\rput[bl](26,0){
%dashed grid
\psset{linewidth=0.25pt,linestyle=dashed, dash=2.5pt 1.5pt,linecolor=gray}
\psline{-}(0,2)(20,2) \psline{-}(0,3)(20,3) \psline{-}(0,4)(20,4)
\psline{-}(0,5)(20,5) \psline{-}(0,6)(20,6) \psline{-}(0,7)(20,7)
\psline{-}(0,1)(0,7)
\psline{-}(1,1)(1,7) \psline{-}(2,1)(2,7) \psline{-}(3,1)(3,7)
\psline{-}(4,1)(4,7) \psline{-}(5,1)(5,7) \psline{-}(6,1)(6,7)
\psline{-}(7,1)(7,7) \psline{-}(8,1)(8,7) \psline{-}(9,1)(9,7)
\psline{-}(10,1)(10,7) \psline{-}(11,1)(11,7) \psline{-}(12,1)(12,7)
\psline{-}(13,1)(13,7) \psline{-}(14,1)(14,7) \psline{-}(15,1)(15,7)
\psline{-}(16,1)(16,7) \psline{-}(17,1)(17,7) \psline{-}(18,1)(18,7)
\psline{-}(19,1)(19,7) \psline{-}(20,1)(20,7)
%axes
\psset{linewidth=0.25pt,fillstyle=none,linestyle=solid,linecolor=black}
\psline{->}(0,1)(21,1)
%\psline{-}(0,9)(22,9)
%numeros
\rput(-0.7,1){\scriptsize $1$}\rput(-0.7,3){\scriptsize $2$}
\rput(-0.7,5){\scriptsize $3$} \rput(-0.7,7){\scriptsize $4$}
%path
\psset{linewidth=0.7pt,fillstyle=none,linestyle=solid,linecolor=black}
\psline(1,1)(7,7)(11,3)(14,6)(19,1)
\psline(11,3)(17,3)
%dots
\psset{dotsize=2.5pt,fillcolor=black}
}
\psline{->}(2,-4.5)(4,-4.5)
\rput[bl](6,-8.5){
%dashed grid
\psset{linewidth=0.25pt,linestyle=dashed, dash=2.5pt 1.5pt,linecolor=gray}
\psline{-}(0,2)(20,2) \psline{-}(0,3)(20,3) \psline{-}(0,4)(20,4)
\psline{-}(0,5)(20,5) \psline{-}(0,6)(20,6) \psline{-}(0,7)(20,7)
\psline{-}(0,1)(0,7)
\psline{-}(1,1)(1,7) \psline{-}(2,1)(2,7) \psline{-}(3,1)(3,7)
\psline{-}(4,1)(4,7) \psline{-}(5,1)(5,7) \psline{-}(6,1)(6,7)
\psline{-}(7,1)(7,7) \psline{-}(8,1)(8,7) \psline{-}(9,1)(9,7)
\psline{-}(10,1)(10,7) \psline{-}(11,1)(11,7) \psline{-}(12,1)(12,7)
\psline{-}(13,1)(13,7) \psline{-}(14,1)(14,7) \psline{-}(15,1)(15,7)
\psline{-}(16,1)(16,7) \psline{-}(17,1)(17,7) \psline{-}(18,1)(18,7)
\psline{-}(19,1)(19,7) \psline{-}(20,1)(20,7)
%axes
\psset{linewidth=0.25pt,fillstyle=none,linestyle=solid,linecolor=black}
\psline{->}(0,1)(21,1)
%\psline{-}(0,9)(22,9)
%numeros
\rput(-0.7,1){\scriptsize $1$}\rput(-0.7,3){\scriptsize $2$}
\rput(-0.7,5){\scriptsize $3$} \rput(-0.7,7){\scriptsize $4$}
%path
\psset{linewidth=0.7pt,fillstyle=none,linestyle=solid,linecolor=black}
\psline(1,1)(6,6)(9,3)(13,7)(19,1)
\psline(3,3)(9,3)
%dots
\psset{dotsize=2.5pt,fillcolor=black}
}
\psline{->}(28,-4.5)(30,-4.5)
\rput[bl](32,-8.5){
%dashed grid
\psset{linewidth=0.25pt,linestyle=dashed, dash=2.5pt 1.5pt,linecolor=gray}
\psline{-}(0,2)(20,2) \psline{-}(0,3)(20,3) \psline{-}(0,4)(20,4)
\psline{-}(0,5)(20,5) \psline{-}(0,6)(20,6) \psline{-}(0,7)(20,7)
\psline{-}(0,1)(0,7)
\psline{-}(1,1)(1,7) \psline{-}(2,1)(2,7) \psline{-}(3,1)(3,7)
\psline{-}(4,1)(4,7) \psline{-}(5,1)(5,7) \psline{-}(6,1)(6,7)
\psline{-}(7,1)(7,7) \psline{-}(8,1)(8,7) \psline{-}(9,1)(9,7)
\psline{-}(10,1)(10,7) \psline{-}(11,1)(11,7) \psline{-}(12,1)(12,7)
\psline{-}(13,1)(13,7) \psline{-}(14,1)(14,7) \psline{-}(15,1)(15,7)
\psline{-}(16,1)(16,7) \psline{-}(17,1)(17,7) \psline{-}(18,1)(18,7)
\psline{-}(19,1)(19,7) \psline{-}(20,1)(20,7)
%axes
\psset{linewidth=0.25pt,fillstyle=none,linestyle=solid,linecolor=black}
\psline{->}(0,1)(21,1)
%\psline{-}(0,9)(22,9)
%numeros
\rput(-0.7,1){\scriptsize $1$}\rput(-0.7,3){\scriptsize $2$}
\rput(-0.7,5){\scriptsize $3$} \rput(-0.7,7){\scriptsize $4$}
%path
\psset{linewidth=0.7pt,fillstyle=none,linestyle=solid,linecolor=black}
\psline(1,1)(4,4)(7,1)(13,7)(19,1)
%dots
\psset{dotsize=2.5pt,fillcolor=black}
}

\end{pspicture}
\end{center}
\end{figure}

If $x>x'$ and $\hh_{x'}=\hh_x+\tfrac12$, then necessarily
$x=x'+2\hat d+\tfrac12$.
The move is then enacted by lowering the peak
at $(x',\hh_{x'})$ by 1/2 unit by removing a NE-SE edge pair,
and heightening the peak at $(x,\hh_x)$ by inserting a NE-SE pair.
This results in peaks at $(x'-\tfrac12,\hh_{x'}-\tfrac12)$ and
$(x-\tfrac12,\hh_x+\tfrac12)$,
with the former now associated with the particle of charge $\hat d$.
This move is illustrated by the transition between the second
and third paths in Fig.~\ref{Fig:Moves2}.
(Note that this latter move may be regarded as two half-moves
of the above types.)

\begin{lemma}\label{Lem:MoveShift}
Let $\hh'$ be obtained from $\hh\in\HH^t_{1,1}$ using one of
the above moves. Then
\begin{equation}\label{Eq:MoveShift}
\wh(\hh')=\wh(\hh)+1.
\end{equation}
\end{lemma}

\Proof
In the first cases, two straight vertices have each moved a
distance of $2\hat d$ to the right, while the $4\hat d-2$ straight
vertices of the particle have each moved a distance of 1
to the left.
In view of \eqref{Eq:WtsDef} and \eqref{Eq:HalfWt}, we therefore
obtain $\wh(\hh')-\wh(\hh)=\tfrac12(4\hat d-(4\hat d-2))=1$.

For the last case, the move may be regarded as having the
$4\hat d-2$ straight vertices of the particle moving to the left
by a distance of $2\hat d+1$, with $4\hat d$ straight vertices of
the other peak moving a distance of $2\hat d$ to the right.
Then, from \eqref{Eq:WtsDef} and \eqref{Eq:HalfWt}, we obtain
$\wh(\hh')-\wh(\hh)=\tfrac12(8\hat d^2-(4\hat d-2)(2\hat d+1))=1$.
\cqfd

Once a particle has moved using the
process described above, it can move again unless its
origin is at the path startpoint $(0,1)$ or lies also
on the slope of a particle of equal or lesser charge.
However, since the particle's origin shifts to the left with each move,
only a finite number of such moves are possible.
For a fixed sector $\boldn$, we define $m_d$ to be the maximal
number of moves that can be performed on the leftmost particle of
charge $d/2$ from $\hh^{\boldn}$ (if there is such a particle).
We now claim that $m_d$, defined in this way,
accords with that specified in Theorem \ref{Thm:HalfFerm}.

\begin{lemma}\label{Lem:MoveCount}
\begin{equation}
\label{Eq:MoveCount}
m_d=\sum_{k=d+1}^{2t-2} n_{k}(k-d).
\end{equation}
\end{lemma}

\Proof
In $\hh^{\boldn}$, the baseline of each particle of charge $d'/2$
has length $d'$.
Therefore, the origin of the leftmost particle of
charge $d$ is at position $(x,1)$ where
$x=\sum_{k=d+1}^{2t-2} kn_{k}$.
Consider repeatedly moving this particle to the left.
Each move of the first type shifts the origin of the particle
1 unit to the left.
However, for each of the particles to its left
(which have greater charge),
one of the other move descriptions must eventually be carried out.
For the second type of move, the change in convention for the
position of the particle, means that the particle's origin shifts
to the left by a distance $d$ before the move is enacted.
For the third type of move, the particle's origin shifts to the
left by a distance $d+1$ in carrying out the move.
Thus, once we have subtracted from $x$ a contribution of $d$ for each
particle of charge greater then $d$, the number of moves, $m_d$,
required to shift the particle's origin to $(0,1)$, remains.
This yields \eqref{Eq:MoveCount}.
\cqfd

\begin{lemma}\label{Lem:SectorFerm}
For $t\in\HZZ$ and
$\boldn=(n_2,n_{3},n_4,\ldots,n_{2t-2})\in\ZZ_{\ge0}^{2t-3}$,
\begin{equation}
\label{Eq:SectorFerm}
\sum_{\hh\in\sector{\boldn}} q^{\wh(\hh)}=
  \frac{q^{\frac12{\boldn}\boldB^{(t)}{\boldn}^T}}{(q)_{m_{1}}}
  \prod_{j=2}^{2t-3}
  \qbinom{n_{j}+m_{j}}{n_{j}},
\end{equation}
where $\boldB^{(t)}$ is defined by \eqref{Eq:Bdef} and
$m_d$ is specified by \eqref{Eq:Mdef} for $1\le d<2t-2$.
\end{lemma}

\Proof
Lemma \ref{Lem:MoveCount} shows that the maximal number of moves for
the leftmost particle of charge $d/2$ in $\hh^{\boldn}$ is given by $m_d$.
If $\lambda^{(d)}_1\le m_d$ moves are actually carried out,
then we may consider moving the next leftmost particle
of charge $d/2$, if there is one.
We see that it may move $\lambda^{(d)}_2\le\lambda^{(d)}_1$ times,
because each of its moves is a shifted version of the corresponding
move of the other particle, and after $\lambda^{(d)}_1$ moves,
it will lie alongside the other, and be unable to move further.
Proceeding likewise with each of the $n_d$ particles of charge $d/2$,
we find that, altogether, their possible moves are indexed by partitions
$\lambda^{(d)}=(\lambda^{(d)}_1,\lambda^{(d)}_2,\ldots,\lambda^{(d)}_{n_d})$,
with largest part at most $m_d$.
Since, by Lemma \ref{Lem:MoveShift}, each move increases the weight of
the path by 1, moving the particles of charge $d/2$ in accordance
with the parts of the partition $\lambda^{(d)}$ increases the weight
by $|\lambda^{(d)}|$, where, as usual, the weight $|\lambda|$ of
a partition $\lambda=(\lambda_1,\lambda_2,\ldots,\lambda_n)$ is
defined by $|\lambda|=\sum_{i=1}^n \lambda_i$.

However, the proof of Lemma \ref{Lem:MoveCount}
applies equally if the particles of charges greater than
$d/2$ have themselves already been subject to moves.
Thus, we may first subject the particles of charge $t-3/2$
to moves indexed by a partition $\lambda^{(2t-3)}$ having at most
$n_{2t-3}$ parts with maximal part $m_{2t-3}$,
followed by subjecting the particles of charge $t-2$
to moves indexed by a partition $\lambda^{(2t-4)}$ having at most
$n_{2t-4}$ parts with maximal part $m_{2t-4}$,
and so on, ending with a partition
$\lambda^{(1)}$ having an unlimited number of parts
(because there are an infinite number of particles of charge 1/2)
but maximal part $m_{1}$.
The weight of a path obtained through this correspondence
is equal to
\begin{equation}
\wh(\hh^{\boldn})+|\lambda^{(2t-3)}|
+|\lambda^{(2t-4)}|+\cdots
+|\lambda^{(1)}|.
\end{equation}
Then, since the generating function for partitions $\lambda$ with at most
$n$ parts with maximal part $m$ is given by \cite{And}
\begin{equation}
\sum_{\begin{subarray}{c} \lambda=(\lambda_1,\lambda_2,\ldots,\lambda_n)\\
                          m\ge\lambda_1\ge\lambda_2\ge\cdots\ge\lambda_n\ge0
       \end{subarray}}
q^{|\lambda|}
=\qbinom{n+m}{n},
\end{equation}
and the generating function for partitions $\lambda$ with an unlimited
number of parts but maximal part $m$ is given by
\begin{equation}
\sum_{\begin{subarray}{c} \lambda=(\lambda_1,\lambda_2,\ldots)\\
                          m\ge\lambda_1\ge\lambda_2\ge\cdots\ge0
       \end{subarray}}
q^{|\lambda|}
=\frac{1}{(q)_m},
\end{equation}
it follows that
the generating function for the set of paths obtained by moving
the particles of $\hh^{\boldn}$ is given by the right side
of \eqref{Eq:SectorFerm}, having made use of Lemma \ref{Lem:MinimalWt}.

It remains to show that every element of $\sector{\boldn}$
is obtained by moving the particles of $\hh^{\boldn}$.
To see this, for an arbitrary path $\hh\in\sector{\boldn}$,
consider first the particles of charge $1/2$ that do
not belong to the tail.
Starting with the rightmost, these may be moved rightward by a
succession of moves reverse to those defined above,
until they belong to the tail.
We then proceed similarly with the particles of charge 1,
starting with the rightmost, and moving it
rightward until it is alongside the tail.
Continuing in this way in turn for all the particles in
$\hh$, we obtain $\hh^{\boldn}$.
By inverting this procedure, it is then clear that $\hh$ can be
obtained from $\hh^{\boldn}$ by moving its particles.
\cqfd

\medskip\noindent
\emph{Proof of Theorem \ref{Thm:HalfFerm}: }
Since every path $\hh\in\HH^t_{1,1}$ is a member of a
unique $\sector{\boldn}$,
we obtain
\begin{equation}
\label{Eq:SumSectorFerm}
Y^t_{1,1}(q)=
\sum_{\boldn\in\ZZ^{2t-3}_{\ge0}}
\sum_{\hh\in\sector{\boldn}} q^{\wh(\hh)}=
  \sum_{\boldn\in\ZZ^{2t-3}_{\ge0}}
  \frac{q^{\frac12{\boldn}\boldB^{(t)}{\boldn}^T}}{(q)_{m_{1}}}
  \prod_{j=2}^{2t-3}
  \qbinom{n_{j}+m_{j}}{n_{j}},
\end{equation}
by summing Lemma \ref{Lem:SectorFerm} over all $\boldn$.
Theorem \ref{Thm:HalfFerm} then follows from Theorem \ref{Thm:HalfChar}.
\cqfd

\section{Discussion}

In this work, we have demonstrated that half-lattice paths
provide a combinatorial model for the $\chi^{p,2p\pm1}_{r,s}$
characters, and give rise to fermionic expressions via
techniques used in the ABF cases.
These expressions are distinct from the fermionic expressions
for these characters obtained previously using the RSOS paths.
The existence of these different fermionic expressions
%$\chi^{p,2p\pm1}_{r,s}$
can be attributed to the fact that there exist different
perturbations of the $\M(p,p')$ conformal field theory 
that remain solvable away from the critical limit \cite{BMcent,BMP}.
The RSOS statistical models \cite{ABF,FB} realise the
so-called $\phi_{1,3}$ perturbation.
We anticipate that the half-lattice paths correspond to
configuration sums of
statistical models that realise the
$\phi_{1,5}$ perturbation when $p'=2p+1$ and the
$\phi_{2,1}$ perturbation when $p'=2p-1$.
The bijections described in this paper would therefore
provide a combinatorial connection between these perturbations
and the $\phi_{1,3}$ perturbation.
However, this connection is not expected to be significant from
the physical point of view.

In a forthcoming paper, we obtain novel fermionic expressions for
all the generating functions $\GH^t_{\hat a,\hat b}(q)$,
and thus, through Theorem \ref{Thm:HalfChar} above,
all characters $\chi^{p,2p\pm1}_{r,s}$.
We also obtain fermionic expressions for the finitizations of these
characters that are the generating functions
of finite length half-lattice paths.
In fact, in each case, we obtain four fermionic expressions.
From a combinatorial point of view, these arise in a way similar to
that of Melzer's expressions \cite{Mel,FWmel} for the ABF cases.

% aaaaaaaaaaaaaaaaaaaaaaaaaaaaaaaaaaaaaaaaaaaaaaaaaaaaaaaaaaaaaaaaaaaaaaa

\begin{appendix}

\section{The Virasoro algebra and the minimal models}

\subsection{The Virasoro algebra}

The Virasoro algebra \vir\ is the infinite-dimensional Lie algebra
over $\CC$ with basis
\begin{equation}
\{\hat c, L_i\,|\,i\in\ZZ\},
\end{equation}
whose elements are subject to the commutation relations
\begin{equation}
\label{Eq:VirCommRels}
\begin{split}
[{L_n},{L_m}]&=
(n-m){L_{n+m}}+\frac{1}{12}n(n^2-1)\delta_{n,-m}{\hat c},\\
[{\hat c},{L_n}]&=0,
\end{split}
\end{equation}
for all $m,n\in\ZZ$.

\subsection{Highest weight modules}
For $c,\Delta\in\CC$,
a highest weight module $V_{c,\Delta}$ of \vir\ is a module of \vir\
that is generated by a vector $v\in V_{c,\Delta}$, for which
\begin{equation}
\label{Eq:HWvector}
\begin{split}
{L_n}\, v&=0 \text{ for all } n>0,\\
{\hat c}\, v&=c\,v,\\
{L_0}\, v&=\Delta\,v.
\end{split}
\end{equation}

The (normalised) character $\chi(V)$ of a highest weight module $V$
is a $q$-series in which the coefficient of $q^n$ is the dimension of
the eigenspace of $V$, on which $L_0$ has eigenvalue $\Delta+n$.
Formally,
\begin{equation}
\chi(V)=q^{-\Delta}\,\text{Tr}_V\, q^{L_0}.
\end{equation}

\subsection{Verma module}
Let $M_{c,\Delta}$ denote the Verma module of \vir\ that has
a highest weight vector $v$ satisfying \eqref{Eq:HWvector}.
The commutation relations in \vir\ imply that
$M_{c,\Delta}$ has a basis comprising all vectors of the form
\begin{equation}
{L_{-n_1}L_{-n_{2}}L_{-n_{3}}\cdots L_{-n_k}}\, v,
\end{equation}
with $n_1\ge n_{2}\ge n_{3}\ge\cdots\ge n_k>0$.

Note that \eqref{Eq:VirCommRels} implies that
\begin{equation}
{L_0}({L_{-n_1}L_{-n_{2}}L_{-n_{3}}\cdots L_{-n_k}}\,v)
=(n+\Delta) ({L_{-n_1}L_{-n_{2}}L_{-n_{3}}\cdots L_{-n_k}}\,v),
\end{equation}
where $n=n_1+n_2+\cdots+n_k$.
It follows that
$\chi(M_{c,\Delta})$ is the ``partition generating function'':
\begin{equation}
\chi(M_{c,\Delta})=\frac{1}{(q)_\infty}.
\end{equation}
%where $(q)_\infty=\prod_{i=1}^\infty(1-q^i)$.

\subsection{Minimal model Virasoro characters}
For generic values of $c$ and $\Delta$,
the Verma module $M_{c,\Delta}$ is irreducible.
The non-generic cases where $c$ is rational are especially
important in physics.
These cases are usually parameterised by four integers
$p,p',r,s$ for which $p$ and $p'$ are coprime with $1<p<p'$,
and $1\le r<p$ and $1\le s<p'$ \cite{KR,DfMS}.
Here
\begin{equation}
\label{Eq:Vir_cDel}
{c=1-6\frac{(p'-p)^2}{pp'}}\qquad\text{and}\qquad
{\Delta=\frac{(p'r-ps)^2-(p'-p)^2}{4pp'}}.
\end{equation}
In these cases, the Verma module $M_{c,\Delta}$
is reducible and its irreducible
quotient has character given by \eqref{Eq:Rocha} \cite{Rocha-Caridi}.
The expression \eqref{Eq:Rocha}
is known as a bosonic expression because it arises
on formulating the Verma module as a bosonic Fock space,
from which submodules are then factored out.
Such expressions are thus the difference of two (genuine) $q$-series.

It is useful to note the following identities which are obtained
directly from \eqref{Eq:Rocha}:
\begin{align}
\label{Eq:RCid1}
\chi^{p,p'}_{r,s}&=\chi^{p,p'}_{p-r,p'-s},\\
\label{Eq:RCid2}
\chi^{p,p'}_{r,s}&=\chi^{p',p}_{s,r}=\chi^{p'/k,kp}_{s/k,kr},
\end{align}
regardless of whether $kp,kr,p'/k,s/k$ are integers.

We will understand here that the ``Minimal Model'' $\M(p,p')$
is the direct sum of all irreducible modules $V_{c,\Delta}$
for $c$ and $\Delta$ obtained from \eqref{Eq:Vir_cDel}
as $r$ and $s$ run over their ranges $1\le r<p$ and $1\le s<p'$.

\end{appendix}


\begin{thebibliography}{99}

\bibitem{And}
G.E. Andrews,
\emph{The Theory of Partitions},
Encyclopedia of Mathematics and its Applications, Vol.~2
(Addison-Wesley, Reading, MA), 1976.

\bibitem{ABF}
G.E. Andrews, R.J. Baxter and P.J. Forrester,
\textit{Eight-vertex SOS model and generalized Rogers-Ramanujan-type
identities},
J. Stat. Phys. \textbf{35} (1984) 193--266.

%\bibitem{AnKi}
%G.E. Andrews and K. Eriksson,
%\emph{Integer Partitions},
%2004 (Cambridge University Press).

%\bibitem{Ba}
%R.J. Baxter,
%\emph{Exactly Solved Models in Statistical Mechanics},
%1982 (Academic Press, London); 2007 (Dover, New York).
 
\bibitem{BM}
A. Berkovich and B.M. McCoy,
\emph{Continued fractions and fermionic representations for
characters of $\M(p,p')$ minimal models},
Lett. Math. Phys. {\bf 37} (1996) 49--66.

\bibitem{BMcent}
A. Berkovich and B.M. McCoy,
\emph{Rogers-Ramanujan identities: A century of progress
from mathematics to physics},
Doc. Math. J. DMV, Extra Volume ICM III (1998) 163--172.

\bibitem{BMS}
A.~Berkovich, B.M.~McCoy and A.~Schilling,
\emph{Rogers-Schur-Ramanujan type identities for the $M(p, p')$ minimal models
  of conformal field theory},
Commun. Math. Phys. {\bf 191} (1998) 325--395.

\bibitem{BMP}
A.~Berkovich, B.M.~McCoy and P.A.~Pearce,
\emph{The perturbations $\phi_{2,1}$ and $\phi_{1,5}$ of the
  minimal models $M(p.p')$ and the trinomial analogue of Bailey's lemma},
Nucl. Phys. {\bf B519} (1998) 597--625.

\bibitem{BfMW}
O.~Blondeau-Fournier, P.~Mathieu and T.A.~Welsh,
\emph{A bijection between paths for the $M(p,2p+1)$ minimal model
      Virasoro characters},
Ann. Henri Poincar\'e {\bf 11} (2010) 101--125.

\bibitem{Bres}
D.M.~Bressoud,
\emph{Lattice paths and the Rogers-Ramanujan identities} in
``Proceedings of the International Ramanujan Centenary Conference,
1987 Madras'',
ed.\ K.~Alladi,
Lecture Notes in Mathematics {\bf1395}
(Springer, New York),
1989,
pp.~140--172.

%\bibitem{bytsko-fring}
%A.G.~Bytsko and A.~Fring,
%\emph{Factorized combinations of Virasoro characters},
%Commun. Math. Phys. {\bf 209} (2000), 179--205.

%\bibitem{CRS}
%J.M. Camino, A.V. Ramallo and J.M. Sanchez de Santos,
%\emph{Graded parafermions},
%Nucl. Phys. {\bf B530} (1998) 715--741.

%\bibitem{christe}
%P.~Christe,
%\emph{Factorized characters and form factors of descendant
%                  operators in perturbed conformal systems},
%Int. J. Mod. Phys. {\bf A6} (1991), 5271--5286.

\bibitem{DJKMO}
E. Date, M. Jimbo, A. Kuniba, T. Miwa and M. Okado,
\emph{Exactly solvable SOS models: local height probabilities
      and theta function identities},
Nucl. Phys. {\bf B290} (1987) 231--273.
 
\bibitem{DfMS}
P.~Di Francesco, P.~Mathieu and D.~S\'en\'echal,
\emph{Conformal Field Theory}
(Springer-Verlag, New York), 1997.

\bibitem{FLPW}
O.~Foda, K.S.M.~Lee, Y.~Pugai and T.A.~Welsh,
\emph{Path generating transforms},
Contemp. Math. {\bf 254} (2000) 157--186.

%\bibitem{FQ}
%O. Foda and Y.-H. Quano,
%\emph{Virasoro character identities from the Andrews-Bailey construction},
%Int. J. Mod. Phys. {\bf A12} (1997) 1651--1675.

\bibitem{FWmel}
O. Foda and T.A. Welsh,
\emph{Melzer's identities revisited},
Contemp. Math. {\bf 248} (1999) 207--234. 

\bibitem{FW-kyoto} % math.QA/0002100
O.~Foda and T.A.~Welsh,
\emph{On the combinatorics of Forrester-Baxter models},
in proceedings of
\lq\lq Physical Combinatorics\rq\rq,
Kyoto 1999, eds. M. Kashiwara and T. Miwa,
Prog. Math. {\bf191}
(Birkh\"auser, Boston),
2000,
pp.~49--103.

\bibitem{FB}
P.J. Forrester and R.J. Baxter,
\emph{Further exact solutions of the eight-vertex SOS model and
generalizations of the Rogers-Ramanujan identities},
J. Stat. Phys. {\bf 38} (1985) 435--472.

%\bibitem{JMgra}
%P. Jacob and P. Mathieu,
%\emph{Graded parafermions: standard and quasi-particle bases},
%Nucl. Phys. {\bf B630} (2002) 433--452.

%\bibitem{JMpath}
%P. Jacob and P. Mathieu,
%\emph{Paths for $\mathcal{Z}_k$ parafermionic models},
%Lett. Math. Phys. {\bf 81} (2007) 211--226.

%\bibitem{PMnlob}
%P. Jacob and P. Mathieu,
%\emph{Nonlocal operator basis from the path representation of
%      the $\M(k+1,k+2)$ and the $\M(k+1,2k+3)$ minimal models},
%J. Phys. A {\bf41} (2008) 385201--21.

\bibitem{PMnpd}
P. Jacob and P. Mathieu,
\emph{A new path description for the ${\mathcal M}(k+1,2k+3)$ models
and the dual $\mathcal{Z}_k$ graded parafermions},
J. Stat. Mech. (2007) P11005 (43 pp).

\bibitem{JMrsos}
P. Jacob and P. Mathieu,
\emph{Particles in RSOS paths},
J. Phys. A {\bf42} (2009) 122001 (14 pp).

\bibitem{KR}
V.~Kac and A.K.~Raina,
\emph{Bombay lectures on highest weight representations of
      infinite dimensional Lie algebras}
(World Scientific, Singapore), 1987.

\bibitem{KKMMferm}
R. Kedem, T.R. Klassen, B.M. McCoy and E. Melzer,
\emph{Fermionic sum representations for conformal field theory characters},
Phys. Lett. B {\bf307} (1993) 68--76.

%\bibitem{PMjmp}
% P. Mathieu,
%\emph{Paths and partitions: combinatorial descriptions of the
%      parafermionic states},
%J. Math. Phys. {\bf 50} (2009) 095210 (43 pp.).

\bibitem{Mel}
E. Melzer,
\emph{Fermionic character sums and the corner transfer matrix},
Int. J. Mod. Phys. {\bf A9} (1994) 1115--1136. 

\bibitem{Rocha-Caridi}
A.~Rocha-Caridi,
\emph{Vacuum vector representations of the Virasoro algebra},
in proceedings of
\lq\lq Vertex Operators in Mathematics and Physics\rq\rq,
eds. J.~Lepowsky et al.
%eds. J.~Lepowsky, S.~Mandelstam and I.M.~Singer,
(Springer-Verlag, New York),
1985,
pp.~451--473.

\bibitem{OleJS1}
S.O.~Warnaar,
\emph{Fermionic solution of the Andrews-Baxter-Forrester model. I.
Unification of CTM and TBA methods},
J. Stat. Phys. \textbf{82} (1996) 657--685.

\bibitem{OleJS2}
S.O.~Warnaar,
\emph{Fermionic solution of the Andrews-Baxter-Forrester model. II.
Proof of Melzer's polynomial identities},
J. Stat. Phys. \textbf{84} (1996), 49--83. 

\bibitem{OleJMP99}
S.O.~Warnaar,
\emph{$q$-Trinomial identities},
J. Math. Phys. {\bf 40} (1999) 2514--2530.

%\bibitem{TWpa}
%T.A.~Welsh,
%\emph{Paths, Virasoro characters and fermionic expressions},
%in proceedings of
%\lq\lq Symmetry and Structural Properties of Condensed Matter\rq\rq,
%Myczkowce, Poland 2005,
%J. Phys.: Conf. Ser. {\bf30} (2006) 119--132.

\bibitem{TWfe}
T.A.~Welsh,
\emph{Fermionic expressions for minimal model Virasoro characters},
Mem. Amer. Math. Soc. {\bf 175} (no.~827) 2005.

\end{thebibliography}
\end{document}